\documentclass[11pt,a4paper]{article}
\pdfoutput=1
\usepackage{jcappub}
\usepackage{hyperref}
\usepackage{subfig}
\hypersetup{colorlinks = true, citecolor = blue}
\usepackage{cancel}

\newcommand{\bvec}[1]{\vec{#1}}

\begin{document}
%
%\preprint{ITP-UU-11/32, SPIN-11/24}
%
\title{Uniqueness of the gauge invariant action
        for cosmological perturbations
       }
\author{Tomislav Prokopec}
\emailAdd{t.prokopec@uu.nl}
\author{and Jan Weenink}
\emailAdd{j.g.weenink@uu.nl}

\affiliation{Institute for Theoretical Physics and Spinoza Institute,\\
Utrecht University, Leuvenlaan 4, 3585 CE Utrecht, The Netherlands}
\abstract{
In second order perturbation theory different definitions are known
of gauge invariant perturbations in single field inflationary models.
Consequently the corresponding gauge
invariant cubic actions do not have the same form. Here we show that
the cubic action for one choice of gauge invariant variables is unique in the
following sense: the action for any other, non-linearly related variable
can be brought to the same bulk action, plus additional boundary terms.
These boundary terms correspond to the choice of hypersurface and generate
extra, disconnected contributions to the bispectrum.
We also discuss uniqueness of the action with respect to conformal frames. When expressed
in terms of the gauge invariant curvature perturbation on uniform field hypersurfaces
the action for cosmological perturbations has a unique form, independent of the original
Einstein or Jordan frame. Crucial is that the gauge invariant comoving curvature
perturbation is frame independent, which makes it extremely helpful in showing
the quantum equivalence of the two frames, and
therefore in calculating quantum effects in nonminimally coupled theories such as Higss inflation.
}
\arxivnumber{}
\maketitle

\section{Introduction}
The detection of non-Gaussianities in the Cosmic Microwave
Background radiation (CMB) would provide a wealth of information.
From the point of view of inflationary physics,
it would constrain the correlation
functions of primordial fluctuations. Since these correlation functions can
be explicitly found for one of the many inflationary theories,
non-Gaussianity provides a powerful tool to constrain the parameters
in these theories and discriminate between models in the inflationary zoo.

One of the simplest inflationary models is that of a single scalar
field in a slow-roll scenario. In his seminal work \cite{Maldacena:2002vr}
Maldacena found the third order action for inflationary
perturbations and showed that non-Gaussianities are too small
to be observed. This has been extended to more general scalar
theories in \cite{Seery:2005wm} and \cite{Chen:2006nt}.

Importantly, the cubic action and 3-point functions have been
derived for $\zeta$, the curvature perturbation. It is well-known
that $\zeta$ itself is a gauge dependent quantity, in the sense
that it is not invariant under reparametrizations of coordinates.
Only a gauge invariant perturbation can be called physical.
On the other hand one can work with gauge fixed quantities,
although one must be careful with respect to gauge artifacts.

In this work we set out to find the cubic action for gauge invariant
cosmological perturbations. We show how to do this for different
gauge invariant variables at second order and that the gauge invariant actions
reduce to the ones in \cite{Maldacena:2002vr} in the spatially flat
or uniform field gauge. In general, the gauge invariant
cubic actions for different variables are different, but we
show that the bulk part of the gauge invariant action coincides for different
variables. In that sense the evolution of non-Gaussianity is unique.
The difference between the actions lies in boundary terms,
which are associated with the choice of hypersurface. They
generate additional, disconnected parts of the bispectrum.

In the second part of this work we discuss uniqueness from
the point of view of conformally related frames. As is well
known, nonminimally coupled actions, or Jordan frame actions,
can be written into a minimally coupled form, the Einstein frame,
by field dependent redefinitions of the metric and scalar field.
Thus in principle nontrivial Jordan frame results can be obtained
from well-known Einstein frame results by redefining fields, which
makes the frame transformation a very powerful tool.
The situation becomes complicated at the level of perturbations.
The problem is that the perturbations of the metric and scalar
field in the Einstein frame are not equal to those in the Jordan
frame, precisely due to the field dependent field redefinition.
The situation is very similar to the gauge problem for perturbations.

In this work we point out that it is possible to construct
\textit{frame independent cosmological perturbations}, which
are very useful to relate Einstein frame results to Jordan frame
results. Here is where the second aspect of \textit{uniqueness}
comes in: the cubic action for cosmological perturbations
takes a unique form, independent of the frame, provided one
makes use of precisely that variable that is the same in either frame.
As it turns out, this variable coincides with the gauge invariant
comoving curvature perturbation.

A short outlook: in Sec. \ref{sec: Action and perturbations}
we define the single scalar field action and its perturbations,
in Sec. \ref{sec: Gauge dependence and gauge invariant perturbations}
we construct different gauge invariant variables at second order
and in Sec. \ref{sec: The gauge invariant action for cosmological perturbations}
we construct the gauge invariant action for cosmological perturbations
and discuss uniqueness. In the second part, Sec.
\ref{sec: Frame independent cosmological perturbations} we discuss
different frames and perturbations in those frames, and show
in what aspect the action is unique.

\section{Action and perturbations}
\label{sec: Action and perturbations}
The action under consideration is the Einstein-Hilbert
action for General Relativity plus a minimally coupled scalar
field
\begin{equation}
S=\int d^4 x \sqrt{-g}\left\{\frac12R
-\frac12 g^{\mu\nu}\partial_{\mu}\Phi\partial_{\nu}\Phi -V(\Phi)\right\}
\,.
\label{EinsteinFrameaction}
\end{equation}
Here we use natural units in which $8\pi G = 1$, $c=1=\hbar$.
This so-called Einstein frame action is manifestly covariant,
that is, it is invariant under spacetime coordinate reparametrizations.
It is possible to study the background field equations for~\eqref{EinsteinFrameaction}.
Taking as the background spacetime the homogeneous and isotropic FLRW metric
leads for example to the Friedmann equations in presence of a time dependent
scalar field, which give rise to inflationary solutions in certain regimes.

It is also possible to study perturbations of the metric and the scalar field
in the action~\eqref{EinsteinFrameaction}. A perturbation of a quantity
is defined as the difference between the quantity in the physical spacetime
and the quantity in the background spacetime. In order to compare
these quantities one has to choose a mapping between the physical and
background spacetimes. The \textit{gauge freedom} in General Relativity is
the freedom in choosing a mapping. As a consequence the perturbations
themselves depend on the choice of mapping.
They are in other words \textit{gauge dependent}.

Gauge dependence is in principle problematic,
since physical results should not depend on reparametrization of coordinates.
However, we know for a fact that the original, unperturbed action~\eqref{EinsteinFrameaction}
is explicitly covariant. Therefore, it may be possible to write the perturbed
action in a manifestly covariant way. Indeed, this can be achieved using
\textit{gauge invariant cosmological perturbations}~\cite{Bardeen:1980kt}.
Gauge invariance here is synonymous to covariance or diffeomorphism invariance.

A convenient method to deal with the gauge dependence in the action
is to use the ADM formalism~\cite{Arnowitt:1962hi} with line element
\begin{equation}
ds^2=-N^2dt^2+g_{ij}(dx^i+N^idt)(dx^j+N^jdt)
\,.
\label{ADMlineelement}
\end{equation}
Geometrically, spacetime has been sliced up
in spatial hypersurfaces whose geometry is described
by the spatial metric $g_{ij}$. The slicing and threading
of spacetime is described by the lapse function $N$
and shift functions $N_i$.
The action~\eqref{EinsteinFrameaction} with the ADM metric~\eqref{ADMlineelement}
becomes
\begin{equation}
S=\frac12\int d^3xdt N \sqrt{g}\Biggl\{R^{(3)}+N^{-2}\left(E^{ij}E_{ij}-E^2\right)\\
+N^{-2}\left(\partial_0{\Phi}-N^{i}\partial_i\Phi\right)^2
-g^{ij}\partial_i\Phi\partial_j\Phi-2V(\Phi)\Biggr\}
\,,
\label{ADMaction}
\end{equation}
where
\begin{align}
\nonumber E_{ij}&=\frac12\left(\partial_0g_{ij}-\nabla_iN_j-\nabla_jN_i\right)\\
E&=g^{ij}E_{ij}
\,,
\end{align}
and $R^{(3)}$ is the spatial scalar curvature computed from
(spatial derivatives of) $g_{ij}$ alone.
From Eq.~\eqref{ADMaction} it is clear that $N$ and $N_i$ are
non-dynamical fields, and moreover they are gauge dependent.
In fact, in a Hamiltonian formulation they
appear as Lagrange multipliers. Solving for these fields in the
action~\eqref{ADMaction} removes the unphysical degrees of freedom.
In the end the only dynamical perturbations out of the 7 degrees
of freedom in $g_{ij}$ and $\Phi$ are one scalar degree of freedom
and a graviton. The remaining 4 degrees of freedom are nondynamical
and are actually the solutions of the constraint equations~\cite{Prokopec:2010be}.
The dynamical degrees of freedom are indeed gauge invariant.

The perturbed action is calculated by inserting
\begin{align}
\nonumber \Phi&=\phi(t)+\varphi\\
\nonumber g_{ij}&=a(t)^2\left(\delta_{ij}e^{2\zeta}+\frac{1}{a^2}\partial_i\partial_j
\tilde{h}+\frac{1}{a}\partial_{\left(i\right.}h^T_{\left.j\right)}+h_{ij}^{TT}\right)\\
\nonumber N&=\bar{N}(t)\left(1+n\right)\\
N_i&=a(t)\bar{N}(t)(\frac{1}{a}\partial_is+n^{T}_i)
\,,
\label{perturbedmetricfield}
\end{align}
where $\zeta$ is the curvature perturbation and
all perturbations carry a temporal and spatial dependence.
We have made use of the scalar-vector-tensor decomposition
of the metric and scalar-vector decomposition of the lapse with
\begin{align}
\partial^{i}h_i^T=0,\qquad\qquad\partial^{i}h_{ij}^{TT}=0=\partial^{j}h_{ij}^{TT},
\qquad\qquad \partial^{i}n^{T}_i=0
\,.
\end{align}
We choose to define every derivative in the decomposition
of the perturbation with a factor of $a$,
such that the combination $a dx$ appears everywhere.

Although apparently linear, the field perturbations contain
in principle perturbations to all orders. For instance,
\begin{equation}
\varphi=\lambda\varphi^{(1)}+\lambda^2\varphi^{(2)}+\mathcal{O}(\lambda^3)
\,,\qquad\qquad
\lambda\ll 1
\,,
\end{equation}
where $\lambda$ indicates the order in perturbations. Similar
expansions hold for the other perturbations in~\eqref{perturbedmetricfield}.
Note that the background value of $N_i$ is zero 
($s,n^{T}_i$ are of $\mathcal{O}(\lambda)$),
since we are describing perturbations on top of a FLRW spacetime.
Moreover, having $\bar{N}(t)$ as a background for $g_{00}$
allows us to rescale time. For example, conformal time would
be defined by setting
$\bar{N}(t)\rightarrow a(\eta)$,
thus having $\bar N(t)dt\rightarrow a(\eta)d\eta$,
with $\eta$ being conformal time.

\section{Gauge dependence and gauge invariant perturbations}
\label{sec: Gauge dependence and gauge invariant perturbations}
As mentioned above the gauge freedom in General Relativity
corresponds to the fact that different mappings can be chosen
between the physical and background spacetime. Changing the mapping
is then referred to as a gauge transformation. If $x^{\mu}$ is the
vector field associated with one mapping, then a different gauge
choice $x^{\mu}+\xi^{\mu}$ transforms a quantity $Q$ according to
\cite{Bruni:1996im,Malik:2008im}
\begin{equation}
Q\rightarrow e^{\mathcal{L}_{\xi}}Q
\,,
\label{general gauge transform}
\end{equation}
where $\mathcal{L}_{\xi}$ is the Lie derivative along $\xi$.
The 4-vector $\xi^{\mu}$ contains all orders in perturbations,
$\xi^{\mu}\equiv \lambda\xi_{(1)}^{\mu}+\frac12 \lambda^2 \xi_{(2)}^{\mu}+\mathcal{O}(\lambda^3)$.
$\xi^{\mu}$ can be separated in a temporal and a spatial part,
which can be written as
\begin{equation}
\xi^{\mu}=\left(\xi^0,\xi^i\right)=\left(\xi^0,\frac{1}{a}\partial^i\xi+\xi^{i(T)}\right)
\,,
\end{equation}
where $\xi^{i(T)}$ is the transverse part of the spatial $\xi^{i}$
($\partial_i \xi^{i(T)}=0$).

\subsection{Gauge transformations of metric and field fluctuations}
We now study more precisely how the scalar and metric field
in \eqref{EinsteinFrameaction} transform under the gauge transformation
\eqref{general gauge transform}. We consider the gauge transformations up to second order in $\lambda$.
For simplicity we only look at gauge transformations of the scalar degrees of
freedom $\varphi$ and $\zeta$, which transform under the temporal gauge
parameter $\xi^0$. We are not interested in second order perturbations of the
other scalars $n$ and $\nabla^2s$,
because they are constraint fields that
can be eliminated from the action by solving the constraint equations.
Alternatively, they can be decoupled from scalar perturbations
in a procedure~\cite{Prokopec:2010be}
that constructs the gauge invariant action including gauge invariant constraints.
Also, we neglect vector and
tensor modes and spatial derivatives which are unimportant on long wavelengths.

For the scalar field $\Phi$ in \eqref{EinsteinFrameaction} the
gauge transformation~\eqref{general gauge transform} 
acts on the perturbation as
\begin{equation}
\varphi\rightarrow \varphi+\dot{\phi} \bar{N}\xi^0
+\frac12(\ddot{\phi} \bar{N}\xi^0+ \dot{\phi}(\bar{N}\xi^0)^{\cdot}
+2\dot{\varphi})\bar{N}\xi^0
+\mathcal{O}(\xi^{i},\partial_i \xi^{0}, \partial_i\varphi)
\,.
\label{scalar gauge transform}
\end{equation}
Here the dotted derivative denotes a reparametrization invariant
time derivative $\dot{\phi}=\frac{d}{\bar{N}dt}\phi$,
such that it is easy to rescale time, for example conformal time by setting
$\bar{N}(t)\rightarrow a(\eta)$. Moreover, the spatial derivatives appear with a factor of
$a$. At higher order
we have only shown quadratic terms containing (temporal derivatives of) $\xi^0$.
Other terms include $\xi^{i}$ and/or spatial derivatives of $\varphi$ or $\xi^0$,
which, as we mentioned, have been neglected.

Similarly, the metric tensor transforms under~\eqref{general gauge transform}.
If we consider the curvature perturbation $\zeta$ then
\begin{equation}
2\zeta\rightarrow  2\zeta
+2H\bar{N}\xi^0+\bar{N}\xi^0\left[
2\dot{\zeta}
+\dot{H}\bar{N}\xi^0
+H(\bar{N}\xi^0)^{\cdot}\right]
+\mathcal{O}(\xi^{i},\partial_i \xi^{0}, \partial_i\zeta)
\,.
\label{zeta gauge transform}
\end{equation}
Here $H\equiv \dot{a}/a$.
Note that $\zeta$ only transforms under temporal gauge transformations
$t\rightarrow t+\xi^0$ at linear order in perturbations. Also here
we have not explicitly written other higher order
terms that include $\xi^{i}$ and/or spatial derivatives of $\zeta$ or $\xi^{0}$.
These terms appear when projecting out the gauge transformation of $\zeta$.
The precise transformations of $\zeta$ and $\varphi$
can be found, for example, in Refs. \cite{Bruni:1996im,Malik:2008im}
or \cite{Noh:2004bc} (the latter does not include the spatial gauge transformation
$\xi^{i}$).

\subsection{Constructing gauge invariant variables}
From Eqs. \eqref{scalar gauge transform} and \eqref{zeta gauge transform} a combination
can be formed which is gauge invariant under temporal gauge transformations to first order.
This combination is
\begin{equation}
w\equiv 2\frac{H}{\dot{\phi}}\varphi-2\zeta
\,,
\label{SasakiMukhanovvariable}
\end{equation}
and is called the comoving curvature perturbation.
This gauge invariant combination of scalar metric
and field perturbations appears not to be unique,
in the sense that it can be rescaled by any function depending
on the background fields ($\dot{\phi}, H$). These
background quantities are by construction fixed
and do not induce additional gauge transformations.
Thus there are in principle infinitely many gauge invariant
combinations. However, a rescaling of a parameter by time dependent
functions does not change any physical results. From
that point of view the first order gauge invariant variable
is unique.

Note that we could have also made a gauge invariant
combination of scalar metric or field perturbations
with the perturbed constraint fields $n$ or $s$ which also
transform under temporal gauge transformations. However,
here we are only interested in dynamical gauge invariant
variables. The constraint fields in the action \eqref{ADMaction}
are nondynamical: they can be solved for and their solutions
inserted back into the action. Thus, the only gauge invariant
dynamical degree of freedom is the comoving curvature perturbation $w$.

Although gauge invariant to
first order in perturbations (governed by the small parameter $\lambda$),
to second order $w$ changes under the gauge transformation
\begin{align}
\nonumber w &\rightarrow w
+\left(\frac{\ddot{\phi}}{H\dot{\phi}}-\frac{\dot{H}}{H^2}\right)
[H^2 (\bar{N}\xi^0)^2+2\zeta H\bar{N}\xi^0]
+\left[\left(\frac{\ddot{\phi}}{H\dot{\phi}}-\frac{\dot{H}}{H^2}\right)H w+\dot{w}\right]\bar{N}\xi^0
+\mathcal{O}(\xi^{i},\partial_i \xi^{0}, \partial_i w)\\
&\equiv w + \Delta_{2,\zeta}^{\xi} w
\,,
\label{second order gauge transform w}
\end{align}
where the second order gauge transformation of $w$
is $\Delta_{2,\zeta}^{\xi} w$, as in Ref. \cite{Rigopoulos:2011eq},
with an additional subscript $\zeta$ which indicates the gauge transformation
involves terms $\propto \zeta$.
$w$ can be made gauge invariant to second order by adding quadratic perturbations
to its definition. For example, we know that to first order $\zeta$ changes under a
gauge transformation as $\zeta\rightarrow \zeta +H\bar{N}\xi^{0}$.
Therefore we can define
\begin{align}
\nonumber W_{\varphi}&=
w-\left[\left(\frac{\ddot{\phi}}{H\dot{\phi}}-\frac{\dot{H}}{H^2}\right)(\zeta^2+w\zeta)
+\frac{1}{H}\dot{w}\zeta\right]\\
&=w-F_{\zeta}[w,\zeta]
\,,
\label{Wzetagaugeinvariant}
\end{align}
which is gauge invariant to second order\footnote{
\label{footnote:GUvariable}Here we have not
written terms with vectors, tensors and spatial derivatives
of $w$ which are unimportant on superhorizon scales. The complete
second order gauge invariant variables can be found in,
for example, \cite{Malik:2008im}.}.
This gauge invariant variable
is related by a factor $H/\dot{\phi}$
to the gauge invariant field perturbation on uniform curvature
hypersurfaces \cite{Malik:2003mv,Malik:2008im},
as one can construct it by fixing the vector field $\xi^{\mu}$
at each order such that $\zeta=0$. Here, we have constructed the
gauge invariant variable by demanding that the gauge transformation at each
order is countered by appropriate terms. We emphasize that we do not want
to use any gauge fixing in this work. Rather, we wish to maintain all fields
and their perturbations, and eliminate any unphysical degrees of freedom
by using gauge invariant variables.

Alternatively we could have countered the temporal gauge dependence in
Eq. \eqref{second order gauge transform w} by quadratic terms in $\varphi$
which transform as $\varphi\rightarrow \varphi + \dot{\phi} \bar{N} \xi^0$.
In that case it is useful to replace $\zeta=\frac{H}{\dot{\phi}}\varphi-\frac12 w$
in the second order gauge transformation of $w$ \eqref{second order gauge transform w}
such that
\begin{align}
\nonumber w &\rightarrow w
+\left(\frac{\ddot{\phi}}{H\dot{\phi}}-\frac{\dot{H}}{H^2}\right)
[H^2 (\bar{N}\xi^0)^2+2\frac{H}{\dot{\phi}}\varphi H\bar{N}\xi^0]
+\dot{w}\bar{N}\xi^0
+\mathcal{O}(\xi^{i},\partial_i \xi^{0}, \partial_i w)\\
&\equiv w + \Delta_{2,\varphi}^{\xi} w
\,.
\label{second order gauge transform w 2}
\end{align}
$\Delta_{2,\varphi}^{\xi} w$ here means the second order gauge transformation
of $w$ involving $\varphi$. Of course the gauge transformation here is the same
as in Eq.~\eqref{second order gauge transform w}, \textit{i.e.}
$\Delta_{2,\varphi}^{\xi} w=\Delta_{2,\zeta}^{\xi} w$, because we have merely
rewritten the gauge transformation in terms of $\varphi$.
Now we can define another variable
\begin{align}
\nonumber W_{\zeta}&=
w-\left[\left(\frac{\ddot{\phi}}{H\dot{\phi}}-\frac{\dot{H}}{H^2}\right)
\frac{H^2}{\dot{\phi}^2}\varphi^2
+\frac{1}{\dot{\phi}}\dot{w}\varphi\right]\\
&=w-F_{\varphi}[w,\varphi]
\,,
\label{Wvarphigaugeinvariant}
\end{align}
which is gauge invariant to second order as well
(see footnote \ref{footnote:GUvariable}). In the
literature \cite{Malik:2003mv,Malik:2008im} this variable is
constructed by fixing the gauge such that $\varphi=0$ at each order,
and is therefore called the curvature perturbation on uniform
field hypersurfaces.

In principle we could have picked any combination of quadratic
perturbations in $\zeta$ and $\varphi$ to balance the second order
gauge dependence of $w$. The advantage of the above variables is
that they reduce to the linear perturbations if one of the scalar perturbations
is set to zero. For example, in the case where $\zeta=0$,
$W_{\varphi}\rightarrow 2\frac{H}{\dot{\phi}}\varphi$, or when $\varphi=0$,
the other gauge invariant variable $W_{\zeta}\rightarrow -2\zeta$.
As it turns out, this makes it very useful to find the gauge invariant
action at third order in terms of these variables. We will discuss this
in the next section.

It is obvious that the two gauge invariant variables $W_{\varphi}$ and
$W_{\zeta}$ are not equal at second order. Their difference can be
expressed in terms of a gauge invariant second order part. For example,
substituting the equality $\zeta=\frac{H}{\dot{\phi}}\varphi-\frac12 w$
in the definition of $W_{\varphi}$ \eqref{Wzetagaugeinvariant}, we find
\begin{align}
\nonumber W_{\varphi}&=W_{\zeta}+\frac{1}{4\dot{\phi}}\left(\frac{\dot{\phi}}{H}\right)^{\cdot}
W_{\zeta}^2+\frac12\frac{1}{H}W_{\zeta}\dot{W}_{\zeta}\\
&=W_{\zeta}+Q(W_{\zeta},W_{\zeta})
\,.
\label{difference gauge invariant variables}
\end{align}
So, the difference between the two gauge invariant variables is
quadratic in $W_{\zeta}^2$ and its derivatives,
and is therefore gauge invariant by itself.
For clarification, at second order $W_{\zeta}^2=W_{\varphi}^2$,
so Eq.~\eqref{difference gauge invariant variables} gives a nonlinear
relation between the different second order gauge invariant variables.
Both gauge invariant variables can be called 'physical' degrees of
freedom, in the sense that they do not depend on the unphysical gauge
degrees of freedom. The question is if the variables describe the same
physics. One could imagine that, since the variables are not equal, the
2-point and 3-point functions are also different and this may give
different results. Of course, in order to calculate the 2-point and
3-point functions and describe the dynamics of the gauge invariant variables,
one should study the action for them, which is what we do next.

\section{The gauge invariant action for cosmological perturbations}
\label{sec: The gauge invariant action for cosmological perturbations}

\subsection{Gauge invariance at zeroth order}
\label{Gauge invariance at zeroth order}
The starting point is the Einstein frame action \eqref{EinsteinFrameaction},
which is manifestly covariant.
Now one can insert the ADM metric \eqref{ADMlineelement} and one
finds -- up to boundary terms -- the action \eqref{ADMaction}. 
Although the general covariance is not
manifest in this action, it is still present. We have merely decomposed
the metric $g_{\mu\nu}$ in separate parts, but as a whole it still transforms
as a tensor.

Now we insert perturbations on top of a fixed homogeneous, isotropic and
expanding background. If we consider the background alone, the action is
\begin{equation}
S^{(0)}=\int d^3xdt\bar{N}a^3\left\{-3H^2+\frac12\dot{\phi}^2-V(\phi)\right\}
.
\label{Einsteinbackgroundaction}
\end{equation}
This action is trivially covariant, in the sense that the background quantities
transform at zeroth order under coordinate transformations, such that the
background fields are fixed functions of the coordinates. In other words, if $a=a(t)$ in one
coordinate system, than $a=a(\tilde{t})$ in another coordinate system.

The hamiltonian constraint, momentum equation and field equation for the background
are found by varying the action with respect to the $\bar{N}$,
$a$ and $\phi$, respectively,
\begin{align}
\nonumber 3H^2&=\frac12 \dot{\phi}^2+V(\phi)\\
\nonumber 2\dot{H}&=-\dot{\phi}^2\\
0&=\ddot{\phi}+3H\dot{\phi}+V'(\phi)
\,.
\label{background field equations}
\end{align}
It turns out that it is useful to define a variable $z$ as
\begin{equation}
z\equiv \frac{\dot{\phi}}{H}
\,,
\label{definition z}
\end{equation}
such that the various slow-roll parameters can be written as,
\begin{align}
\nonumber \epsilon &\equiv -\frac{\dot{H}}{H^2}=\frac12 z^2\\
\eta &\equiv \frac{\dot{\epsilon}}{\epsilon H}=2\frac{\dot{z}}{zH}
\,.
\label{definition slow roll parameters}
\end{align}
Here the same definitions have been used as in \cite{Seery:2005wm}.
These slow-roll parameters are very useful for finding the dominant
contributions to $n$ point functions from the action.

\subsection{Gauge invariance of quadratic action}
\label{sec: Gauge invariance at second order}
The action to linear order in perturbations vanishes
due to the classical background equations of motion.
The first nontrivial action of perturbations is
the second order action in perturbations. Usually
one first eliminates the constraint fields $N$ and $N_i$
by solving for them and inserting their solutions back
into the action. It is only necessary to do this to
first order in perturbations since the second order
solutions multiply the classical equations of motion.
If we consider only the scalar fluctuations $\zeta$ and
$\varphi$, the resulting second order action looks 
schematically like
\begin{equation}
S^{(2)}(\zeta,\varphi)=\int d^3xdt\bar{N}a^3
\left\{A\mathcal{O}(\zeta^2)+B\mathcal{O}(\zeta\varphi)
+C\mathcal{O}(\varphi^2)\right\}
\,,
\label{second order action approximate}
\end{equation}
where $\zeta$ and $\varphi$ are now linear in the perturbation
parameter $\lambda$, such that the action is of order $\lambda^2$.
With the indication $\mathcal{O}(\zeta^2)$ we mean all terms of order
$\zeta^2$, including $\dot{\zeta}^2$, $\zeta\dot{\zeta}$,
$\partial_i\zeta\partial^{i}\zeta$ and possible mixings. A similar
reasoning applies to the $\mathcal{O}(\varphi^2)$ and $\mathcal{O}(\zeta\varphi)$ terms.
The explicit form \eqref{second order action approximate} 
can be found in \textit{e.g.} \cite{Mukhanov:1990me} or \cite{Prokopec:2010be}.

As was mentioned in Sec.~\ref{sec: Action and perturbations}
(and shown explicitly in Sec.~\ref{sec: Gauge dependence and gauge invariant perturbations}),
the perturbations $\zeta$ and $\varphi$ transform under linear gauge transformations,
specifically under temporal gauge transformations. Therefore, the variables
$\zeta$ and $\varphi$ do not separately have a physical meaning, as they
depend on our choice of coordinate system. The combination $w$ in Eq.
\eqref{SasakiMukhanovvariable} is on the other hand gauge invariant,
and if we express the schematic action~\eqref{second order action approximate}
in terms of this $w$, one obtains~\cite{Mukhanov:1990me}
\begin{equation}
S^{(2)}(w)=\int d^3x dt\bar{N}a^3 \frac14 z^2
\left\{\frac12 \dot{w}^2-\frac12 \left(\frac{\partial_i w}{a}\right)^2\right\}
\,,
\label{Mukhanov gauge invariant action}
\end{equation}
up to total derivative terms. $z$ was defined in Eq.~\eqref{definition z}.
Apart from $w$, the action also contains the
transverse, traceless metric perturbation $h_{ij}$, in short the graviton,
which is automatically gauge invariant to first order.\footnote{Instead of solving for the constraint
fields, it is also possible to keep the constraint fields and perturb them
to second order \cite{Prokopec:2010be}. After decoupling them from the dynamical degrees
of freedom there are an additional 4 non-dynamical degrees of freedom in the
second order action, which are gauge invariant as well.
Varying the action with respect to these degrees of freedom
gives the first order solutions of the constraint fields. Thus, out of 11 degrees
of freedom in the metric and scalar field, we are left with 3 dynamical and 4 constraint
degrees of freedom. The 4 gauge degrees of freedom have thus been eliminated from the action.}

The action \eqref{Mukhanov gauge invariant action} is now manifestly gauge invariant
up to second order in perturbations: the variables are diffeomorphism invariant
up to a first order in coordinate reparametrizations, and can therefore be considered
physical, and the background fields are trivially gauge invariant. One can now do
the usual steps of rescaling $w$ to a variable $v=\frac12 a^{\frac32}\frac{\dot{\phi}}{H}w$,
the Mukhanov variable, such that the action becomes that of a harmonic oscillator
with a time dependent mass. One can then quantize $v$ with the usual expansion in
creation and annihilation operators, solve for the mode functions in, for example,
an inflationary background, and calculate the power spectrum for those perturbations
that remain constant on superhorizon scales (which coincides with the comoving curvature
perturbation $w$ on superhorizon scales during slow-roll inflation).

\subsection{Gauge invariance of cubic action}
\label{sec:Gauge invariance at third order}
In order to study non-Gaussianities in the Cosmic Microwave Background radiation (CMB),
$n$-point functions such as the primordial bispectrum or trispectrum must be derived.
The $3$-point functions are found from the third order action in cosmological
perturbations. Of course, the physical bispectrum is only found when the perturbations
in the (tree-level) cubic action are physical, \textit{i.e} gauge invariant.

Schematically, the third order action in $\zeta$ and $\varphi$ takes the form
\begin{equation}
S^{(3)}(\zeta,\varphi)=\int d^3xdt\bar{N}a^3
\left\{A\mathcal{O}(\zeta^3)+B\mathcal{O}(\zeta^2\varphi)
+C\mathcal{O}(\zeta\varphi^2)+D\mathcal{O}(\varphi^3)\right\}
\,.
\label{third order action approximate}
\end{equation}
Again, one can find this action by dropping vector and tensor perturbations
and solving the constraint equations to first order in perturbations.
The third order solutions multiply the background equations of motion,
and the second order solutions multiply the Hamiltonian and momentum
constraint evaluated at first order. The $A\mathcal{O}(\zeta^3)$ and $D\mathcal{O}(\varphi^3)$
terms were first derived in Ref.~\cite{Maldacena:2002vr}, and for completeness
we give the terms here explicitly. First the $A\mathcal{O}(\zeta^3)$ terms
up to temporal and spatial boundary terms:
\begin{align}
\nonumber S^{(3)}(\zeta)=&\int d^3x dt\bar{N}a^3\Biggl\{
-(\zeta^2+2\zeta\frac{\dot{\zeta}}{H})\frac{\nabla^2\zeta}{a^2}
-(\zeta+\frac{\dot{\zeta}}{H})\left(\frac{\partial_i\zeta}{a}\right)^2
+\frac{3}{2}z^2 \zeta\dot{\zeta}^2
-\frac12 \frac{z^2}{H}\dot{\zeta}^3\\
&+\frac12 \left(3\zeta-\frac{\dot{\zeta}}{H}\right)
\left[\frac{\partial_i\partial_j\psi}{a^2}\frac{\partial_i\partial_j\psi}{a^2}
-\left(\frac{\nabla^2\psi}{a^2}\right)^2\right]
-2\frac{\partial_i\psi}{a}\frac{\partial_i\zeta}{a}\frac{\nabla^2\psi}{a^2}
\Biggr\}
\,,
\label{Cubic Action Zeta}
\end{align}
where
\begin{equation}
\frac{\nabla^2\psi}{a^2}=-\frac{\nabla^2}{a^2}\frac{\zeta}{H}+\frac12 z^2\dot{\zeta}
\,.
\end{equation}
Secondly the $D\mathcal{O}(\varphi^3)$ terms:
\begin{align}
\nonumber S^{(3)}(\varphi)=&\int d^3x dt\bar{N} a^3 \Biggl\{
-\frac14 z \dot{\varphi}^2\varphi
-\frac14 z \left(\frac{\partial_i \varphi}{a}\right)^2\varphi
-\dot{\varphi}\frac{\partial_i\chi}{a}\frac{\partial_i \varphi}{a}\\
\nonumber &+\frac{1}{4}z\left[\varphi\left(\frac{\nabla^2\chi}{a^2}\right)^2
-\varphi\left(\frac{\partial_i\partial_j\chi}{a^2}\right)
\left(\frac{\partial_i\partial_j\chi}{a^2}\right)\right]\\
&+\left[\frac18 H^2 z^3(3-\frac12 z^2)
-\frac14 z V''-\frac16  V'''\right]\varphi^3
+\frac14 z^3 H \varphi^2\dot{\varphi}+\frac14 z^2 \varphi^2 \frac{\nabla^2\chi}{a^2}
\Biggr\}
\,,
\label{Cubic Action varphi}
\end{align}
where
\begin{equation}
\frac{\nabla^2\chi}{a^2}=\frac12 z^2\left(-\frac{\varphi}{z}\right)^{\cdot}
\,.
\end{equation}
The $A\mathcal{O}(\zeta^3)$ terms have been derived for generalized scalar theories
in Ref.~\cite{Seery:2005wm}, see also \cite{Chen:2006nt}.

\subsubsection{Manifest gauge invariance: cubic action for $W_{\varphi}$}
The action \eqref{third order action approximate}
does not appear to be covariant due to the gauge dependence of $\zeta$
and $\varphi$. However, we know that the complete action is covariant, and should
be covariant to this order as well. To make this more manifest, we can try to
express the cubic action in terms of the linear gauge invariant variable $w$,
since the third order terms in~\eqref{third order action approximate} only transform
under linear gauge transformations. In doing so we have several possibilities.

The first option is to eliminate all $\varphi$ dependence in the action by replacing
\begin{equation}
\varphi=\frac12 z w + z\zeta
\,,
\label{varphi replacement}
\end{equation}
where again $z=\dot{\phi}/H$ such that
\begin{equation}
S^{(3)}=\int d^3xdt\bar{N}a^3
\left\{D\mathcal{O}\left(\left(\frac12 z w\right)^3\right)
+E\mathcal{O}(\zeta^3)+F\mathcal{O}(\zeta^2 w)
+G\mathcal{O}(\zeta w^2)\right\}=S_{\rm{GI}}^{(3)}(w)+S^{(3)}_{\rm{GD}}(\zeta)
\,.
\label{third order action approximate in w}
\end{equation}
The first part $S_{\rm{GI}}^{(3)}(w)$ now only depends on $\mathcal{O}(w^3)$ terms,
and is therefore explicitly gauge invariant at third order in $\lambda$.
The second part $S^{(3)}_{\rm{GD}}(\zeta)$ is on the other hand gauge dependent,
because it contains cubic terms $\zeta^3$, $\zeta^2 w$ and $\zeta w^2$ that
cannot balance each other's gauge transformations.
However, the original action~\eqref{EinsteinFrameaction} is diffeomorphism invariant,
and this gauge dependence should somehow cancel. One should not forget
that also the second order action changes under a gauge transformation. Although
\eqref{Mukhanov gauge invariant action} is gauge invariant to first order
in $\lambda$, to second order in gauge transformations it is not.
The gauge transformation~\eqref{second order gauge transform w}
thus generates third order terms in $\lambda$
from the second order action~\eqref{Mukhanov gauge invariant action}.
Following Rigopoulos~\cite{Rigopoulos:2011eq},
under a second order gauge transformation of $w$,
\begin{equation}
S^{(2)}(w)\rightarrow S^{(2)}(w)+\int d^3x dt \bar{N} a^3
\left\{\frac{1}{a^3}\frac{\delta S^{(2)}(w)}{\delta{w}}\Delta_{2,\zeta}^{\xi} w\right\}
\,,
\label{gauge transform second order action}
\end{equation}
where $\Delta_{2,\zeta}^{\xi} w$ was defined in \eqref{second order gauge transform w}.
In order to keep the general covariance at each order in perturbation theory,
this second order gauge transformation of $S^{(2)}(w)$ must be balanced by
appropriate linear gauge transformations of $S^{(3)}$. This means that the
gauge dependent terms in Eq.~\eqref{third order action approximate in w}
must be proportional to the first order equations of motion, such that
\begin{equation}
S^{(3)}_{\rm{GD}}(\zeta)=-\int d^3x dt \bar{N} a^3
\left\{\frac{1}{a^3}\frac{\delta S^{(2)}(w)}{\delta{w}} F_{\zeta}[w,\zeta]\right\}
\,,
\label{third order action propto eom}
\end{equation}
where $F_{\zeta}[w,\zeta]$ is defined in Eq.~\eqref{Wzetagaugeinvariant}
\footnote{This implies that the terms of $\mathcal{O}(\zeta^3)$ should drop out
in the third order action \eqref{third order action approximate}
after the replacement of $\varphi$ \eqref{varphi replacement}.}.
Of course, the gauge dependent part $S^{(3)}_{\rm{GD}}(\zeta)$ is precisely
such that it counters the second order gauge transformation of $S^{(2)}(w)$
in \eqref{gauge transform second order action}.
Now it is straightforward to express the action in terms of the second
order gauge invariant variable $W_{\varphi}$,
\begin{equation}
S^{(2)}+S^{(3)}=\int d^3x dt\bar{N}a^3 \frac14 z^2
\left\{\frac12 \dot{W}_{\varphi}^2-\frac12 \left(\frac{\partial_i W_{\varphi}}{a}\right)^2\right\}
+ S_{\rm{GI}}^{(3)}(W_{\varphi})
\,.
\label{Gauge Invariant Action Wzeta}
\end{equation}
Thus, we have found the third order action which is manifestly gauge invariant
up to second order in gauge transformations. The dynamical scalar degree of
freedom in this action is $W_{\varphi}$, where $\frac12 z W_{\varphi}$ is
the field perturbation on uniform curvature
hypersurfaces. We haven't mentioned explicitly what is $S^{(3)}(W_{\varphi})$,
that is, what are the gauge invariant cubic vertices for $W_{\varphi}$ needed for the
calculations of non-Gaussianities. The easiest way to find this is to set $\zeta=0$
from the start. In that case $W_{\varphi}$ coincides
with $\frac{2}{z}\varphi$ and
\begin{equation}
S_{\rm{GI}}^{(3)}(W_{\varphi})\xrightarrow{\zeta\rightarrow 0}
\int d^3xdt\bar{N}a^3
\left\{D\mathcal{O}\left(\varphi^3\right)\right\}
\,.
\end{equation}
Hence, $S_{\rm{GI}}^{(3)}(W_{\varphi})$ can be found
immediately after the replacement $\varphi\rightarrow \frac12 z W_{\varphi}$
in the $D\mathcal{O}(\varphi^3)$ terms of \eqref{third order action approximate}
\cite{Rigopoulos:2011eq}. Using these terms from~\cite{Maldacena:2002vr} the result is
\begin{align}
\nonumber S_{\rm{GI}}^{(3)}(W_{\varphi})=&\int d^3x dt \frac{\bar{N} a^3}{8} \Biggl\{
\frac{1}{2}z^4\left[-\frac12 \dot{W}_{\varphi}^2W_{\varphi}
-\frac12\left(\frac{\partial_i W_{\varphi}}{a}\right)^2W_{\varphi}
+\dot{W}_{\varphi}\left(\frac{\partial_i}{\nabla^2} \dot{W}_{\varphi}\right)\partial_i W_{\varphi}\right]\\
\nonumber &+\frac{1}{16}z^6\left[\dot{W}_{\varphi}^2W_{\varphi}
-\left(\frac{\partial_i\partial_j}{\nabla^2}\dot{W}_{\varphi}\right)
\left(\frac{\partial_i\partial_j}{\nabla^2}\dot{W}_{\varphi}\right)W_{\varphi}\right]\\
\nonumber &+\left[\frac18 H z^6-\frac12 z^3 \dot{z}\right] W_{\varphi}^2\dot{W}_{\varphi}
+\frac12 z^3\dot{z}W_{\varphi}\left(\frac{\partial_i}{\nabla^2} \dot{W}_{\varphi}\right)\partial_i W_{\varphi}\\
&+\left[-\frac14 z^2\dot{z}^2+\frac14 H z^5 \dot{z}+\frac18 H^2 z^6(3-\frac12 z^2)
-\frac14 z^4 V''-\frac16 z^3 V'''\right]W_{\varphi}^3
\Biggr\}
\,.
\label{Cubic GI Action Wzeta1}
\end{align}
This demonstrates the usefulness of the field perturbation on uniform
curvature hypersurfaces $\frac12 z W_{\varphi}$ in combination with the uniform curvature gauge
$\zeta=0$. The cubic action \eqref{Cubic GI Action Wzeta1} can be simplified
further by doing partial integrations of the third and last lines and using
the background equations of motion \eqref{background field equations},
\begin{align}
\nonumber S_{\rm{GI}}^{(3)}(W_{\varphi})=&\int d^3x dt \frac{\bar{N} a^3}{8} \Biggl\{
\frac{1}{2}z^4\left[-\frac12 \dot{W}_{\varphi}^2W_{\varphi}
-\frac12\left(\frac{\partial_i W_{\varphi}}{a}\right)^2W_{\varphi}
+\dot{W}_{\varphi}\left(\frac{\partial_i}{\nabla^2} \dot{W}_{\varphi}\right)\partial_i W_{\varphi}\right]\\
&+\frac{1}{16}z^6\left[\dot{W}_{\varphi}^2W_{\varphi}
-\left(\frac{\partial_i\partial_j}{\nabla^2}\dot{W}_{\varphi}\right)
\left(\frac{\partial_i\partial_j}{\nabla^2}\dot{W}_{\varphi}\right)W_{\varphi}\right]
-\frac12 z^2 \dot{W}_{\varphi}W_{\varphi}^2
\left[\frac{\dot{z}}{zH}\right]^{\cdot}
\Biggr\}
\,.
\label{Cubic GI Action Wzeta}
\end{align}
In Eq. \eqref{Cubic GI Action Wzeta}
different orders in slow-roll can easily be distinguished. Using the slow-roll
parameters \eqref{definition slow roll parameters} it is clear that the first
line contains terms of order $\epsilon^2$, whereas the second line is subleading in slow-roll.

In the derivation of the gauge invariant action for $W_{\varphi}$ we have set $\zeta=0$.
What happens if we would have taken $\varphi=0$ from the start? In that case
the third order action \eqref{third order action approximate}
contains only terms $A\mathcal{O}(\zeta^3)$, and $W_{\varphi}$
is a nonlinear expression in $\zeta$. Therefore the third order action must contain
terms proportional to the first order equation of motion for $\zeta$, which can be
absorbed in the second order action by identifying $W_{\varphi}(\varphi=0)$. These terms
proportional to the equation of motion were identified in~\cite{Maldacena:2002vr}.
It was shown that after the field redefinition $\zeta\rightarrow \frac12 W_{\varphi}(\varphi=0)$,
the cubic action for $\varphi$ is obtained by replacing
$W_{\varphi}\rightarrow\frac{2H}{\dot{\phi}}\varphi$. In our language of gauge invariance,
we redefine $\zeta$ to the gauge invariant variable $W_{\varphi}$ (in the gauge $\varphi=0$),
then restore the dependence on $\varphi$ in $W_{\varphi}$, and finally set $\zeta=0$ to
get the action for $\mathcal{O}(\varphi^3)$ terms.

\subsubsection{Manifest gauge invariance: cubic action for $W_{\zeta}$}

In the previous section we found the manifestly gauge invariant cubic action
in terms of $W_{\varphi}$, related to the field perturbation on uniform
curvature hypersurfaces. Just as well we could have constructed the gauge invariant action
for $W_{\zeta}$, the curvature perturbation on uniform field hypersurfaces.
Starting point is again the schematic third order action~\eqref{third order action approximate}
and we follow the same steps as in the previous part. Instead of eliminating $\varphi$
in terms of $w$ and $\zeta$ as in Eq.~\eqref{varphi replacement}, we eliminate $\zeta$
\begin{equation}
\zeta=\frac{1}{z}\varphi-\frac12 w
\,.
\label{Zeta replacement}
\end{equation}
As before, the third order action $S^{(3)}$ can then be separated
in a gauge invariant part depending only on $\mathcal{O}(w^3)$ terms,
and a gauge non-invariant part depending on $\varphi$. Of course, this
gauge dependent part must again balance the gauge transformation of $S^{(2)}$
\eqref{gauge transform second order action}. Therefore
\begin{align}
\nonumber S^{(3)}&=\int d^3xdt\bar{N}a^3
\left\{A\mathcal{O}\left(\left(-\frac12w\right)^3\right)
+E'\mathcal{O}(\varphi^3)+F'\mathcal{O}(\varphi^2 w)
+G'\mathcal{O}(\varphi w^2)\right\}\\
\nonumber &=\tilde{S}_{\rm{GI}}^{(3)}(w)+\tilde{S}^{(3)}_{\rm{GD}}(\varphi)\\
&=\tilde{S}_{\rm{GI}}^{(3)}(w)-\int d^3x dt \bar{N} a^3
\left\{\frac{1}{a^3}\frac{\delta S^{(2)}(w)}{\delta{w}} F_{\varphi}[w,\varphi]\right\}
\,.
\label{third order action approximate in w 2}
\end{align}
Again we can do a redefinition of $w$ to the second order
gauge invariant $W_{\zeta}$ using~\eqref{Wvarphigaugeinvariant}
such that
\begin{equation}
S^{(2)}+S^{(3)}=\int d^3x dt\bar{N}a^3 \frac14 z^2
\left\{\frac12 \dot{W}_{\zeta}^2-\frac12 \left(\frac{\partial_i W_{\zeta}}{a}\right)^2\right\}
+ \tilde{S}_{\rm{GI}}^{(3)}(W_{\zeta})
\,.
\label{Gauge Invariant Action Wvarphi}
\end{equation}
Thus, following this procedure one obtains
the manifestly gauge invariant action at third order
expressed in terms of the gauge invariant variable $W_{\zeta}$. The
simplest way to find the gauge invariant vertices in
$\tilde{S}_{\rm{GI}}^{(3)}(W_{\zeta})$ is now to set the field
perturbation $\varphi=0$. Then, the gauge invariant variable
becomes $W_{\zeta}(\varphi=0)=-2\zeta$, and
\begin{equation}
\tilde{S}_{\rm{GI}}^{(3)}(W_{\zeta})\xrightarrow{\varphi\rightarrow 0}
\int d^3xdt\bar{N}a^3
\left\{A\mathcal{O}\left(\zeta^3\right)\right\}
\,.
\end{equation}
So, if we set $\varphi=0$, the third order action for $\zeta$ immediately
gives the gauge invariant action in terms of the curvature perturbation
on uniform field hypersurfaces after the replacement $\zeta\rightarrow -\frac12 W_{\zeta}$.
Using the $A\mathcal{O}(\zeta^3)$ terms derived in~\cite{Maldacena:2002vr,Seery:2005wm},
\begin{align}
\nonumber \tilde{S}_{\rm{GI}}^{(3)}(W_{\zeta})=&\int d^3x dt\frac{\bar{N}a^3}{8}\Biggl\{
\frac12 z^2 W_{\zeta}\left(\frac{\partial_i W_{\zeta}}{a}\right)^2
-\frac{3}{2}z^2 W_{\zeta}\dot{W}_{\zeta}^2
+\frac12 \frac{z^2}{H}\dot{W}_{\zeta}^3\\
&-\frac12 \left(3W_{\zeta}-\frac{\dot{W}_{\zeta}}{H}\right)
\left[\frac{\partial_i\partial_j\psi}{a^2}\frac{\partial_i\partial_j\psi}{a^2}
-\left(\frac{\nabla^2\psi}{a^2}\right)^2\right]
+2\frac{\partial_i\psi}{a}\frac{\partial_iW_{\zeta}}{a}\frac{\nabla^2\psi}{a^2}
\Biggr\}
\,,
\label{Cubic GI Action Wvarphi}
\end{align}
where
\begin{equation}
\frac{\nabla^2\psi}{a^2}=-\frac{\nabla^2}{a^2}\frac{W_{\zeta}}{H}+\frac12 z^2\dot{W}_{\zeta}
\,.
\end{equation}
This demonstrates the convenience of working with the gauge invariant
variable $W_{\zeta}$ in combination with the gauge $\varphi=0$
\footnote{In the derivation of \eqref{Cubic GI Action Wvarphi}
we have performed several partial integrations with respect to
spatial derivatives. The corresponding boundary terms do not contribute
to the bispectrum, contrary to the boundary terms for temporal
partial integrations, which will be discussed next.}.

\section{Uniqueness of gauge invariant action}
\label{sec: Uniqueness of gauge invariant action}
In the previous section two third order actions
for cosmological perturbations were derived which
were manifestly gauge invariant up to second order
in gauge transformations. The general trick is that
the gauge dependent parts of the third order action
could be absorbed in the second order action, which
defined a gauge invariant variable. In
Eq.~\eqref{Gauge Invariant Action Wzeta} the gauge
invariant cubic action was expressed in terms of
$W_{\varphi}$, in Eq.~\eqref{Gauge Invariant Action Wvarphi}
in terms of $W_{\zeta}$. Comparing the gauge invariant
actions \eqref{Gauge Invariant Action Wzeta} and
\eqref{Gauge Invariant Action Wvarphi}, we see that the
parts quadratic in the gauge invariant variables are the same.
Thus the tree-level propagator for $W_{\varphi}$ is the same
for $W_{\zeta}$. The gauge invariant parts of the action
which are cubic in the gauge invariant variables, Eqs.
\eqref{Cubic GI Action Wzeta} and \eqref{Cubic GI Action Wvarphi},
appear not to be the same,
\begin{equation}
S_{\rm{GI}}^{(3)}(W_{\varphi})\neq \tilde{S}_{\rm{GI}}^{(3)}(W_{\zeta})
\,.
\label{Difference gauge invariant actions}
\end{equation}
This implies that the gauge invariant vertices
for $W_{\varphi}$ differ from those for $W_{\zeta}$. On the other hand,
the gauge invariant
cubic vertices originate from the same action. This presents an
opportunity to find out exactly how the gauge invariant actions differ.

\subsection{Non-linear transformations}
To illustrate what is the difference between gauge invariant actions
for non-linearly related variables, let us take a general action
for a second order gauge invariant variable $W_X$,
\begin{equation}
S(W_X)=\int d^3x dt\bar{N}a^3 \frac14 z^2
\left\{\frac12 \dot{W}_{X}^2-\frac12 \left(\frac{\partial_i W_{X}}{a}\right)^2\right\}
+ S_{\rm{GI}}^{(3)}(W_{X})
\,.
\label{uniq:actionWX}
\end{equation}
This variable is non-linearly related to another second order gauge invariant variable
$W_Y$,
\begin{equation}
W_X=W_Y+Q(W_Y,W_Y)
\,,
\label{uniq:relationWXWY}
\end{equation}
where the $Q(W_Y,W_Y)$ is a completely general function quadratic in $W_Y$, which can
include temporal and/or spatial derivatives of $W_Y$. The action for $W_{Y}$ then
becomes
\begin{align}
\nonumber S(W_Y)&=\int d^3x dt\bar{N}a^3 \frac14 z^2
\left\{\frac12 \dot{W}_{Y}^2-\frac12 \left(\frac{\partial_i W_{Y}}{a}\right)^2\right\}
+ S_{\rm{GI}}^{(3)}(W_{Y})\\
&+\int d^3x dt\bar{N}a^3\left\{ \frac{1}{a^3}\left[\frac14 a^3 z^2 Q \dot{W}_Y \right]^{\cdot}
-\frac{1}{a}\partial_i\left[Q\frac{1}{a}\partial_i W_{Y} \right] \right\}
+\int d^3x dt\bar{N}a^3\left\{ Q \frac{1}{a^3}\frac{\delta S^{(2)}(W_{Y})}{\delta{W_{Y}}} \right\}
\,,
\label{uniq:actionWY}
\end{align}
where $\delta S^{(2)}/\delta{W_Y}$ represents the equation
of motion following from the quadratic action, \textit{i.e.}
\begin{equation}
\frac{1}{a^3}\frac{\delta S^{(2)}(W_{Y})}{\delta{W_{Y}}}
=\left[-\frac14 a^3 z^2 \dot{W}_Y\right]^{\cdot}+\frac{1}{a^2}\nabla^2 W_{Y}
\,.
\label{uniq:eomWY}
\end{equation}
These terms are related to the variation of the complete action as
\begin{equation}
\frac{1}{a^3}\frac{\delta S(W_{Y})}{\delta{W_{Y}}}
=\frac{1}{a^3}\frac{\delta S^{(2)}(W_{Y})}{\delta{W_{Y}}}+\mathcal{O}(W_{Y}^2)=0
\,,
\label{uniq:eomAction}
\end{equation}
which vanishes by the variational principle.
In the third order action
the terms $\propto Q \delta S^{(2)}/\delta{W_{Y}}$ therefore are
zero at the cubic level upon inserting the solutions of the
equations of motion~(\ref{uniq:eomAction}),
and consequently these terms do not contribute
to the 3-point function at tree-level. However, contributions
enter in the tree-level action at higher order that
can contribute to the 2- and 3-point function through quantum loops,
which we do not consider here.

Inspecting the tree-level cubic gauge invariant action for $W_Y$ \eqref{uniq:actionWY}
we see that the bulk action $S_{\rm{GI}}^{(3)}(W_{Y})$
coincides with the gauge invariant cubic action for $W_X$ \eqref{uniq:actionWX}.
In this sense the bulk gauge invariant cubic action can be called \textit{unique}.
Thus, the evolution of the 3-point function is independent of
the choice variables $W_X$ or any nonlinearly related variable,
characterized by $W_Y$.

Of course there are also boundary terms in the action for $W_Y$
\eqref{uniq:actionWY}. In the in-in or Schwinger-Keldysh formalism
they can contribute to the 3-point function. In this formalism
an expectation value may be defined as
\begin{align}
\nonumber \langle \Omega, t_{\rm in} | \mathcal{O}(W(t)) | \Omega, t_{\rm in} \rangle
=\int & [\mathcal{D}W_{+}\mathcal{D} W_{-}]\mathcal{O}(W(t))
\rho[W_+(t_{\rm in}),W_-(t_{\rm in})]\\
&\times\exp\left(i\int_{t_{\rm in}}^{t}dt'\left(L[W_{+}(t')]-L[W_{-}(t')]\right)\right)
\delta\left[W_{+}(t)-W_{-}(t)\right]
\,,
\label{uniq:ininexpvalue}
\end{align}
where $\rho[W_+(t_{\rm in}),W_-(t_{\rm in})]$ is the density matrix at initial time $t_{\rm in}$,
which for a pure initial state equals $\Psi^*[W_+(t_{\rm in})]\Psi[W_-(t_{\rm in})]$.
$W_\pm$ is here a second order gauge invariant variable on the $+$ or $-$ part of the
complex {\it in-in} contour, which
could be, for instance, $W_X$ or $W_Y$. In general, the operator $\mathcal{O}(W(t))$
depends on both $W_+$ and $W_-$ fields. In the simple case
of an equal-time 3-point function, operator ordering is not important, one can drop
the $\pm$ subscripts from $W$'s and
$\mathcal{O}(W(t))=
W(\vec{x}_1,t) W(\vec{x}_2,t) W(\vec{x}_3,t)$.
Coming back to the boundary terms, the spatial ones do not contribute
to the 3-point function. On the other hand, the temporal boundary
terms give in general a nonzero contribution.\\
For cosmological correlation functions~\eqref{uniq:ininexpvalue}
the initial time is often taken at $t_0=-\infty$.
Strictly speaking, this is not the correct procedure.
As $t\rightarrow -\infty$ the quantum field theory of gravity
becomes strongly coupled (that is, in that limit the physical momenta
${k}/{a}\rightarrow \infty$) and perturbation theory fails.
In practice one can define an in-in expectation value by starting from some
finite initial time $t_0$ at which perturbation theory
is well defined. Thus boundary terms at $t_0$ can contribute.
One can define the initial state for, for instance, the
gauge invariant variable $W_X$ to be Gaussian. As a consequence
the initial state for any other non-linearly related variable $W_Y$
is explicitly non-Gaussian. This initial non-Gaussianity
is then evolved through the bulk action $S_{\rm{GI}}^{(3)}(W_{Y})$.
It is important to distinguish how much non-Gaussianity is dynamically
generated from some Gaussian initial state, and how much comes from
a potentially non-Gaussian initial state. For example, if one observes
non-Gaussianity for the variable $W_Y$, but one defines the initial
state for $W_X$ to be Gaussian, then some of the final non-Gaussianity
finds its origin in the initial non-Gaussianity in $W_Y$.

Let us now discuss the contributions to the 3-point function coming from the
temporal boundary terms at time $t$. There can be various types of
boundary terms. Boundary terms of the type $W^3(t)$ cannot
contribute to the 3-point function, because the $\delta$-function
in Eq.~\eqref{uniq:ininexpvalue} forces the "$+$" and "$-$" vertices
to be equal at time $t$. In fact, these boundary terms do not
naturally appear after the transformation to a different gauge
invariant variable, as can be seen in \eqref{uniq:actionWY}.
Nonetheless these terms can appear after additional partial integrations
of terms $\propto W^3(t)$ and $\propto W^2(t)\dot{W}(t)$ in the bulk action.
This was exactly done for the action of $W_{\varphi}$ in going from
Eq.~\eqref{Cubic GI Action Wzeta1} to \eqref{Cubic GI Action Wzeta},
which justifies the use of \eqref{Cubic GI Action Wzeta} as the gauge
invariant cubic action for $W_{\varphi}$.\\
One has to be careful with other boundary terms such as
$\alpha W^2\dot{W}$ or $\beta W\dot{W}^2$, because they give in general
non-negligible contributions to the bispectrum. The reason is that
these terms contain the canonical momentum $\Pi_W$, which has
a nonvanishing commutation relation with $W$. In fact these type of
boundary terms generate disconnected parts of the bispectrum. In
the general example above, it can be seen that the temporal boundary terms
in the action for $W_Y$ are precisely of the form
$S_{\partial}(W_Y)=\int d^3 x \Pi_{W_Y}(t) Q(t)$. Consider now an
example where
\begin{equation}
Q(W_Y,W_Y)=\alpha(t)W_Y^2 + \beta(t) W_Y \dot{W}_Y
\,.
\end{equation}
Then using the expectation value as defined in
Eq.~\eqref{uniq:ininexpvalue} one can compute
to lowest (tree-level) order that\footnote{An alternative
way to compute the 3-point correlator~\eqref{uniq:bispectrumWXWY}
is by making use of the so-called interaction picture~\cite{Maldacena:2002vr,Weinberg:2005vy},
where an expectation value to lowest order in perturbation theory
is given by
\begin{equation}
\langle \mathcal{O}(W(t)) \rangle =
-i \int_{t_0}^{t}dt'\langle \left[\mathcal{O}(W(t)),H_{\rm int}(t')\right]\rangle
\,.
\end{equation}
If for the interaction Hamiltonian $H_{\rm int}$ one considers the part
with the boundary terms in~\eqref{uniq:actionWY} only, which are of the form
$\Pi_W(t) Q(t)$, then it is straightforward to find the disconnected pieces
in \eqref{uniq:bispectrumWXWY} using the canonical commutation relations.
}
\begin{align}
\nonumber \langle W_{X}(x_1) W_{X}(x_2)
W_{X}(x_3)\rangle &= \langle W_{Y}(x_1)
W_{Y}(x_2)W_{Y}(x_3)\rangle\\
\nonumber &+ 2\alpha(t) \left(
\langle W_{Y}(x_1)W_{Y}(x_2)\rangle
\langle W_{Y}(x_1)W_{Y}(x_3)\rangle
+\rm{sym}\right)\\
\nonumber &+ \beta(t) \biggl(
\langle \dot{W}_{Y}(x_1)W_{Y}(x_2)\rangle
\langle W_{Y}(x_1)W_{Y}(x_3)\rangle\\
&+\langle W_{Y}(x_1)W_{Y}(x_2)\rangle
\langle \dot{W}_{Y}(x_1)W_{Y}(x_3)\rangle
+\rm{sym}\biggr)
\,,
\label{uniq:bispectrumWXWY}
\end{align}
where sym stands for other cyclic contributions.
In words: the 3-point function for the variable $W_X$
is computed using the gauge invariant cubic vertices
in $S_{\rm{GI}}^{(3)}(W_{X})$, the result expressed
on the left-hand side, but it is also directly
related to the 3-point function for $W_Y$, computed
using gauge invariant vertices for $W_Y$, plus
additional disconnected parts coming from the boundary
terms, which add up to the right-hand side of \eqref{uniq:bispectrumWXWY}.
Note that the relation between the 3-point functions
in \eqref{uniq:bispectrumWXWY} can be immediately derived
by inserting the non-linear relation between $W_X$ and $W_Y$
\eqref{uniq:relationWXWY} into the left-hand side of
\eqref{uniq:bispectrumWXWY} and using Wick's theorem.
Thus in order to compute the 3-point function of one
gauge invariant variable, one can use the gauge invariant
action for another variable (which may have a more convenient form)
and add disconnected pieces according to the non-linear relation.

Since time $t$ in~(\ref{uniq:bispectrumWXWY}) is arbitrary,
relation~\eqref{uniq:bispectrumWXWY} holds also for $t=t_{\rm in}$,
telling us how are initial non-Gaussianities in the variables
$W_X$ and $W_Y$ related. These variables also define some spacelike
hypersurfaces $\Sigma_{W_X}$ and $\Sigma_{W_Y}$.
For example, if the initial state on $\Sigma_{W_Y}$ is Gaussian,
then the initial non-Gaussianity on $\Sigma_{W_Y}$ will be given by
the terms multiplying $\alpha$ and $\beta$ in~(\ref{uniq:bispectrumWXWY})
evaluated at $t=t_{\rm in}$.

\subsection{Practical example: different gauge invariant variables}
The general discussion in this section demonstrates that
the cubic gauge invariant actions \eqref{uniq:actionWX}--
\eqref{uniq:actionWY}
for different~\footnote{Different here means non-linearly
related. This in turn means that the gauge invariant variables
coincide at the linear level, but differ at quadratic order.}
gauge invariant variables are related, in the sense that they both
have the same, unique, bulk action, but they differ by boundary
terms and terms proportional to the equation of motion.
This is not always obvious. For example, the terms
proportional to the equation of motion can be separated,
partially integrated, and the remaining terms can be written
such that it is not clear that they are total derivative terms.
Therefore gauge invariant actions for different, non-linearly
related variables can appear very different. To illustrate this,
let us now consider a practical example. The gauge invariant
variables $W_{\varphi}$ and $W_{\zeta}$
are non-linearly related as in~\eqref{difference gauge invariant variables},
which is an example of \eqref{uniq:relationWXWY}.
According to the above discussion, their 3-point functions
should therefore be related
\begin{align}
\nonumber \langle W_{\varphi}(x_1) W_{\varphi}(x_2)
W_{\varphi}(x_3)\rangle &= \langle W_{\zeta}(x_1)
W_{\zeta}(x_2)W_{\zeta}(x_3)\rangle\\
&+ \frac12\frac{\dot{z}}{zH} \left(
\langle W_{\zeta}(x_1)W_{\zeta}(x_2)\rangle
\langle W_{\zeta}(x_1)W_{\zeta}(x_3)\rangle
+\rm{sym}\right)
+\ldots
\,,
\label{uniq:bispectrumWvarphi}
\end{align}
where the terms of higher order in slow-roll have been neglected.
Note that this relation can also be inverted to give the
3-point function of $W_{\zeta}$ in terms $W_{\varphi}$.
This is purely based on the non-linear relation between the
gauge invariant variables, but it should follow from the actions as well.
The gauge invariant actions for $W_{\varphi}$ and $W_{\zeta}$ coincide
at the quadratic level, but the cubic actions Eq.~\eqref{Cubic GI Action Wzeta}
and \eqref{Cubic GI Action Wvarphi} look very different
at first sight. For instance, the action for $W_{\varphi}$~\eqref{Cubic GI Action Wzeta}
is of second order in slow-roll ($\epsilon^2=\frac14 z^4$), whereas
the action for $W_{\zeta}$ \eqref{Cubic GI Action Wvarphi} seems to
be of zeroth order. Maldacena~\cite{Maldacena:2002vr}
showed that it is possible to relate the cubic action for $\varphi$ in the $\zeta=0$ gauge
with the cubic action for $\zeta$ in the $\varphi=0$ gauge. The two actions differ
by terms proportional to the linear equation of motion 
(which do not contribute to the tree-level action)
and by some boundary terms. When translated to our language of gauge invariant variables,
the cubic action for $W_{\varphi}$ (defined on the $\zeta=0$ hypersurface) 
can be related to $W_{\zeta}$ (defined on the $\varphi=0$ hypersurface), 
up to boundary terms. After many partial integrations of \eqref{Cubic GI Action Wvarphi}, 
the result is:
\begin{align}
\nonumber \tilde{S}_{\rm{GI}}^{(3)}(W_{\zeta})=&\int d^3x dt \frac{\bar{N} a^3}{8} \Biggl\{
\frac{1}{2}z^4\left[-\frac12 \dot{W}_{\zeta}^2W_{\zeta}
-\frac12\left(\frac{\partial_i W_{\zeta}}{a}\right)^2W_{\zeta}
+\dot{W}_{\zeta}\left(\frac{\partial_i}{\nabla^2} \dot{W}_{\zeta}\right)\partial_i W_{\zeta}\right]\\
\nonumber &+\frac{1}{16}z^6\left[\dot{W}_{\zeta}^2W_{\zeta}
-\left(\frac{\partial_i\partial_j}{\nabla^2}\dot{W}_{\zeta}\right)
\left(\frac{\partial_i\partial_j}{\nabla^2}\dot{W}_{\zeta}\right)W_{\zeta}\right]
-\frac12 z^2 \dot{W}_{\zeta}W_{\zeta}^2
\left[\frac{\dot{z}}{zH}\right]^{\cdot}\\
\nonumber &+Q(W_{\zeta},W_{\zeta})\frac{1}{a^3}\frac{\delta S^{(2)}(W_{\zeta})}{\delta{W_{\zeta}}}
\Biggr\}\\
=&~S_{\rm{GI}}^{(3)}(W_{\zeta})+
\int d^3x dt \frac{\bar{N} a^3}{8}
\Biggl\{
Q(W_{\zeta},W_{\zeta})\frac{1}{a^3}\frac{\delta S^{(2)}(W_{\zeta})}{\delta{W_{\zeta}}}
\Biggr\}
\,.
\label{Cubic GI Action Wvarphi after integrations}
\end{align}
This is precisely the form of the action predicted after insertion of
the non-relation relation $W_{\varphi}=W_{\zeta}+Q(W_{\zeta},W_{\zeta})$,
as in \eqref{uniq:actionWY}, up to boundary terms. The gauge invariant
vertices in the bulk action $S_{\rm{GI}}^{(3)}$ for $W_{\zeta}$ coincide
with those for $W_{\varphi}$. Thus,
the cubic gauge invariant action for $W_{\varphi}$ can be called the
unique action which separates the different levels of slow-roll.
A gauge invariant action in terms
of any other gauge invariant variable, \textit{e.g.} $W_{\zeta}$
can be brought to this unique
form after many partial integrations and extracting terms
proportional to the equation of motion. The bispectrum for $W_{\zeta}$ can now
be computed by making use of the partially integrated bulk action,
including possible contributions coming from boundary terms
which can give rise to disconnected contributions to 3-point
functions, see Eq. \eqref{uniq:bispectrumWXWY}--\eqref{uniq:bispectrumWvarphi}.
Alternatively, one can redefine the field $W_{\zeta}$
to the non-linearly related $W_{\varphi}$ and use the action
for that gauge invariant variable. The 3-point function for $W_{\zeta}$
is then computed from Eq. \eqref{uniq:bispectrumWvarphi}.
The non-linear relation between
different gauge invariant variables prescribes what
this field redefinition should be.

\subsection{Boundary terms, hypersurfaces and observations}
So far we have not shown the boundary terms
in \eqref{Cubic GI Action Wvarphi after integrations}.
They were explicitly computed for $\zeta$ (or $W_{\zeta}$
in $\varphi=0$ gauge) in Refs. \cite{Arroja:2011yj}
and \cite{Burrage:2011hd} for scalar field Lagrangians which
are a general function $\Phi$ and its kinetic term. Both
reach same conclusions: boundary terms with time derivatives
of $W_{\zeta}$ contribute to the bispectrum, and the
dominant terms in the slow-roll approximation give exactly the
same contribution as in Eq.~ \eqref{uniq:bispectrumWvarphi}, which
is what one finds after a "field redefinition" to a non-linear
variable in $W_{\zeta}$.\footnote{In fact, in Ref. \cite{Burrage:2011hd}
the procedure is slightly different than stated above. After partial integrations
Burrage et al. do not keep any boundary terms proportional $W_{\zeta}^2\dot{W}_{\zeta}$.
The reason is precisely that these would generate disconnected parts of the
3-point function. The partial integrations performed are only the "allowed" ones:
those that do not contribute to the bispectrum at all, or those that are slow-roll
suppressed contributions. The field redefinition is slightly different in their work,
and it does not coincide with the variable $W_{\varphi}$ (note: all computations
are performed in the comoving, $\varphi=0$, gauge.)}
In our language this is nothing more than
switching between different gauge invariant variables.

One remark here is that the boundary terms in Refs. \cite{Arroja:2011yj}
and \cite{Burrage:2011hd} do not disappear completely after
redefining $W_{\zeta}$ to a new non-linear variable. On the other
hand, Eq. \eqref{uniq:actionWY} suggests that all boundary
terms are incorporated after switching to a non-linear variable.
This must be so because under a non-linear transformation
3-point functions of different variables are related as
\eqref{uniq:bispectrumWXWY}, irrespective of a specific action.
When describing now different gauge invariant variables
with their corresponding actions, the same relation of the
bispectra should follow from the action, for both the bulk
and the boundary.
The origin of these additional boundary terms not removed
by the field redefinition may reside in additional boundary terms of the
quadratic action. The form of the
quadratic action in \eqref{Mukhanov gauge invariant action} is only
reached after several partial integrations, which generate additional
second order boundary terms. Moreover, the original, unperturbed
ADM action \eqref{ADMaction} also contains several spatial
and temporal boundary terms that in principle contribute at every
order. Together these boundary terms must add up
to a gauge invariant second order boundary term, expressed in the linear
$w$, since after all the original starting point is the covariant action
\eqref{EinsteinFrameaction}. Now, under a non-linear field transformation
these boundary terms will generate also cubic boundary terms, which
may balance the extra boundary terms mentioned before.

A different way to see this is to come back to the procedure
of finding the gauge invariant cubic action, outlined in Sec.
\ref{sec:Gauge invariance at third order}. Here the non-linear
gauge transform of the second order \textit{bulk} action was balanced by
gauge dependent terms in the third order action, which are
proportional to the linear equation of motion. This in turn defined
a second order gauge invariant variable ($W_{\varphi}$ or $W_{\zeta}$
depending on the procedure). Similarly, also the second order
boundary terms transform under non-linear gauge transformations.
They can be written in a gauge invariant way after incorporating
the gauge dependent third order boundary terms.

In spite of these remarks
the conclusion of \cite{Arroja:2011yj} and \cite{Burrage:2011hd}
still stands: the dominant contribution to the bispectrum coming
from the boundary terms for $W_{\zeta}$
is taken into account by switching to a different gauge invariant
variable $W_{\varphi}$ and using \eqref{uniq:bispectrumWvarphi}.
This can be very useful. Ultimately we are interested
in describing correlation functions of $W_{\zeta}$ (the second
order gauge invariant comoving curvature perturbation), because this is the field
that is conserved on super-horizon scales and forms the initial
fluctuations in the gravitational potential that are believed
to form the seeds of structure formation and the observed temperature
fluctuations in the CMB. The correct gauge invariant cubic action
to use is therefore Eq. \eqref{Cubic GI Action Wvarphi}. However,
this action does not clearly separate the dominant contributions
in the slow-roll approximation. Before horizon crossing it is therefore
much more useful to work with the non-linearly related variable
$W_{\varphi}$, for which it is straightforward to determine the
dominant vertices. Note that one has to be careful concerning the initial state:
if the initial state for $W_{\zeta}$ is Gaussian at $t_0$, then
for $W_{\varphi}$ it will be automatically non-Gaussian.
After horizon crossing the full action for $W_{\zeta}$ should be used,
as this variable is conserved on superhorizon scales.

Finally some remarks about non-Gaussianity and observations.
As argued, the variable to use to calculate the primordial
power spectrum and non-Gaussianity is $W_{\zeta}$.
The CMB power spectrum, and possible non-Gaussianities there,
are namely created by
fluctuations in the gravitational potential. The universe
reheats at a $\varphi=0$ (more generally $\delta\rho=0$) hypersurface
where photons decouple. These photons carry
the information of the gravitational fluctuations towards us.
However, we observe these
photons via satellites such as WMAP or Planck, that are
to a good approximation freely-falling observers for which
$\zeta=0$, \textit{i.e.} observations are made on zero curvature
hypersurfaces. Choosing a certain gauge invariant variable to work
with, is in essence nothing else but choosing the hypersurface.
\footnote{Albeit the choice of a hypersurface seemingly relies on a gauge variant concept
of setting {\it e.g.} $\varphi=0$ (comoving gauge), or $\zeta=0$
(zero curvature gauge), these surfaces have a well defined physical meaning in
the following sense. Namely, choosing {\it e.g.} $\varphi=0$ completely fixes a gauge
such that in that gauge $-2\zeta$ can be associated with
$W_\zeta$, and hence has a gauge invariant meaning. In that sense we can
talk about a gauge invariant choice of hypersurfaces, $\Sigma_{W_\zeta}$
and $\Sigma_{W_\varphi}$ being just examples of two commonly used hypersurfaces.
In this language, the question on which hypersurface one should perform calculations
of non-Gaussianity becomes immaterial, as long as one knows on which hypersurface
the observer measures, and how to relate the gauge invariant variables associated with
the two hypersurfaces.}
We have seen that different gauge invariant variables are
non-linearly related, and therefore their $n$-point functions
are related via disconnected pieces. As a consequence,
on one hypersurface perturbations may appear non-Gaussian, whereas
they are in fact Gaussian on a different hypersurface, or vice versa.
It is important to distinguish the amount of non-Gaussianity that
is generated by the cosmological evolution, and
non-Gaussianity that originates from the choice of hypersurface, either
initially or at time of observation.

\section{Frame independent cosmological perturbations}
\label{sec: Frame independent cosmological perturbations}
In the first part of this work we have discussed uniqueness
of the gauge invariant action for cosmological perturbations
with respect to the choice of gauge invariant variables.
In this part we consider a different type of uniqueness
for the action for cosmological perturbations, namely
uniqueness with respect to different conformal frames.
Examples are the Einstein frame, Jordan frame or string
frame, which are related via field dependent conformal
transformations of the metric and redefinitions of
the scalar field. Our goal here is to show that at the
level of perturbations the (cubic) action can be written in
a unique form, which is independent of the Einstein or Jordan
frame.

\subsection{Einstein frame and Jordan frame}
\label{sec: Einstein frame and Jordan frame}
So far we have discussed perturbations of the Einstein
frame action~\eqref{EinsteinFrameaction}. It is also possible to study
perturbations in the Jordan frame, where a non-minimal coupling between
the scalar field and the Ricci scalar is present. The Jordan frame action
is
\begin{equation}
S_J=\int d^4 x \sqrt{-g_J}\left\{\frac12 R_J F(\Phi_J)
-\frac12 g_J^{\mu\nu}\partial_{\mu}\Phi_J\partial_{\nu}\Phi_J -V_J(\Phi_J)\right\}
\,.
\label{JordanFrameAction}
\end{equation}
All subscripts $J$ indicate that the quantities are expressed in the Jordan frame.
The function $F(\Phi_J)$ presents the general coupling between the Ricci scalar and
the scalar field $\Phi_J$. Setting $F(\Phi_J)=1(=m_{pl}^2\equiv 1/(8\pi G))$
takes us back to the minimally coupled case. An example of a model with nonminimal coupling
is Higgs inflation~\cite{Salopek:1988qh,Bezrukov:2007ep} where $F(\Phi_J)=1+\xi\Phi_J^2$,
$\xi\gg 1$, and $\Phi_J$ is the Higgs field.

The Einstein frame and the Jordan frame are related via a combined
conformal transformation of the metric and a redefinition of the scalar field
\begin{align}
\nonumber g_{\mu\nu}&=\Omega^2 g_{\mu\nu,J}\\
\nonumber
\left(\frac{d\Phi}{d\Phi_J}\right)^2
&=\frac{1}{\Omega^2}+6\frac{\Omega^{\prime 2}}{\Omega^2}\\
V(\Phi)&=\frac{1}{\Omega^4}V_J(\Phi_J)
\,,
\label{EinsteinJordanFrameRelations}
\end{align}
with
\begin{equation}
\Omega^2=\Omega^2(\Phi_J)=F(\Phi_J)
\,.
\end{equation}
Since these are just field redefinitions of the metric
and scalar field, no physical information
is expected to be lost in the frame transformation.
This is what we refer to as \textit{physical equivalence of
Jordan and Einstein frame}.
The physical equivalence is very useful, because it
means we could obtain any results, such as the power spectrum,
in the Jordan frame by transforming the well-known Einstein frame
results using the above relations~\eqref{EinsteinJordanFrameRelations}.
Instead of dealing with the difficult nonminimal coupling,
we merely have to deal with a modified potential.

Although the physical equivalence between Jordan and Einstein frame,
in the sense described above,
is well established at the classical level, it is not
obvious that it also holds at the level of quantum fluctuations.
In order to solve this issue, one could derive the tree-level
action for cosmological perturbations in the Jordan frame. As a first
step, we again slice up our space-time using the ADM metric
\eqref{ADMlineelement} and insert perturbations similar
to Eqs.~\eqref{perturbedmetricfield},
\begin{align}
\nonumber g_{ij,J}&=a_J(t)^2\left(\delta_{ij}+h_{ij,J}(t,\bvec{x})\right)\\
\nonumber \Phi_J&=\phi_J(t)+\varphi_J(t,\bvec{x})\\
\nonumber N_J&=\bar{N}_J(t)\left(1+n_J(t,\bvec{x})\right)\\
N_{i,J}&=a_J(t)\bar{N}_J(t)n_{i,J}(t,\bvec{x})
\,.
\label{perturbedmetricfieldJordan}
\end{align}
The background Jordan frame action
\eqref{JordanFrameAction} then becomes
\begin{equation}
S_J^{(0)}=\int d^3xdt\bar{N}_Ja_J^3
\left\{-3H_J^2F-3H_J\dot{F}+\frac12\dot{\phi}_J^2-V_J(\phi_J)\right\}
\,,
\label{Jordanbackgroundaction}
\end{equation}
Here $H_J=\frac{\dot{a}_J}{a_J}=\frac{da_J/(\bar{N}_Jdt)}{a_J}$ and
$\dot{\phi_J}=d\phi_J/(\bar{N}_J dt)$.\footnote{A dotted derivative on
a Jordan frame quantity implies that it is a derivative with respect to
$\bar{N}_J dt$, whereas a dotted derivative on an Einstein frame quantity
implies it is with respect to $\bar{N}dt$. We will use this consistently throughout
this work.} Moreover, $F=F(\phi_j)$, so
$F$ only depends on the background field $\phi_J$.
The background equations in the Jordan frame are now obtained
by varying the action with respect to $\bar{N}_J$, $a_J$ and $\phi_J$,
\begin{align}
\nonumber 3H_J^2F&=\frac12 \dot{\phi}_J^2+V_J(\phi_J)-3H_J\dot{F}\\
\nonumber 2\dot{H}_JF&=-\dot{\phi}_J^2+H_J\dot{F}-\ddot{F}\\
0&=\ddot{\phi}_J+3H_J\dot{\phi}_J-3(2H_J^2+\dot{H}_J)F'+V_J'(\phi_J)
\,.
\label{Jordan background equations}
\end{align}
Using Eqs.~\eqref{EinsteinJordanFrameRelations} we can find the
explicit relations between background quantities in the Jordan
and Einstein frame,
\begin{align}
\nonumber \bar{N}&=\bar{\Omega}\bar{N}_J\\
\nonumber a&=\bar{\Omega}a_J\\
\nonumber H&=\frac{1}{\bar{\Omega}}\left(H_J+\frac{\dot{\bar{\Omega}}}{\bar{\Omega}}\right)\\
\dot{\phi}&=\frac{1}{\bar{\Omega}}\frac{d\phi}{d\phi_J}\dot{\phi}_J
\equiv\frac{1}{\bar{\Omega}}
\sqrt{\frac{1}{\bar{\Omega}^2}+6\frac{\bar{\Omega}^{\prime 2}}{\bar{\Omega}^2}}\dot{\phi}_J
\,,
\label{EinsteinJordan Background Relations}
\end{align}
where $\bar{\Omega}^2\equiv \Omega^2(\phi_J)=F(\phi_J)$.
If we substitute these relations into the background action in the Einstein
frame~\eqref{EinsteinFrameaction} or into the field equations~\eqref{background field equations},
we immediately recover the background Jordan frame action~\eqref{Jordanbackgroundaction}
and its background field equations~\eqref{Jordan background equations}.
This establishes the classical equivalence of Jordan and Einstein frame.

\subsection{Frame independence of second order action}
\label{sec: Frame independence of second order action}
Next step is to find the second order action for cosmological perturbations
in the Jordan frame by inserting \eqref{perturbedmetricfieldJordan} into
\eqref{JordanFrameAction}. The nonminimal coupling between metric and
scalar field is making this procedure much more complicated compared
to the Einstein frame. It would be useful to establish also the frame equivalence
at the level of perturbations, such that we can use the Einstein frame results
and transform to the Jordan frame.

Before we proceed, we notice that the metric and field
perturbations are not the same in Jordan and Einstein frame. The conformal
factor $\Omega^2=\Omega^2(\Phi_J)$ can be perturbed to second order as
\begin{equation}
\Omega(\Phi_J)=\bar{\Omega}+\Omega'\varphi_J
+\frac12 \Omega^{\prime\prime}\varphi_J^2
\,,
\label{Omega perturbation}
\end{equation}
where $\Omega'\equiv d\Omega/d\phi_J$.
Using this and the general relations between Jordan and Einstein
frame~\eqref{EinsteinJordanFrameRelations}, we can write
second order relations for $\varphi$ and $\zeta$
\begin{align}
\nonumber \varphi&=\frac{d\phi}{d\phi_J}\varphi_J
+\frac12 \frac{d^2\phi}{d\phi_J^2}\varphi_J^2\\
\zeta&=\zeta_J+\frac{\Omega'}{\bar{\Omega}}\varphi_J
+\frac12 \frac{\Omega^{\prime\prime}}{\bar{\Omega}}\varphi_J^2
-\frac12 \left( \frac{\Omega'}{\bar{\Omega}}\right)^2\varphi_J^2
\,.
\label{JordanEinstein perturbation relation}
\end{align}
Obviously the Jordan and Einstein frame perturbations of the scalar field
and the scalar part of the metric are not the same. Consequently, neither
$\varphi_J$ nor $\zeta_J$ are convenient variables to establish
the quantum equivalence between the two frames, as these variables
are inherently frame dependent. However, from the Einstein frame analysis
we know that neither $\varphi$ nor $\zeta$ are physical perturbations, because
they are gauge dependent. In the second order action for cosmological perturbations
\eqref{Mukhanov gauge invariant action}, we have seen that the only physical
scalar perturbation is $w$, defined in~\eqref{SasakiMukhanovvariable}. With
the Jordan-Einstein frame relations for the
background~\eqref{EinsteinJordan Background Relations} and
perturbations~\eqref{JordanEinstein perturbation relation} at hand it is
straightforward to show that at first order
\begin{equation}
w=2\frac{H}{\dot{\phi}}\varphi-2\zeta=2\frac{H_J}{\dot{\phi}_J}\varphi_J-2\zeta_J \equiv w_J
\,.
\end{equation}
The gauge invariant $w$ in Einstein frame coincides with the gauge invariant $w_J$
in the Jordan frame to first order. Thus the physical comoving curvature perturbation
is not only \textit{gauge invariant}, but \textit{frame independent} as well.
This frame independence to first order was first proven in
Refs.~\cite{Makino:1991sg,Fakir:1992cg}. Using this
one can immediately write down the second order Jordan frame action
via a frame transformation from the second order
Einstein frame action~\eqref{Mukhanov gauge invariant action},
which was done in, for instance, Ref. \cite{Hwang:1996np}.
By making use of the relation
\begin{equation}
z^2=\frac{\dot{\phi}^2}{H^2}=\frac{1}{\bar{\Omega}^2}
\frac{\dot{\phi}_J^2+6\dot{\bar{\Omega}}^2}{(H_J+\frac{\dot{\bar{\Omega}}}{\bar{\Omega}})^2}
\equiv \frac{1}{\bar{\Omega}^2} z_J^2
\,,
\label{definition zj}
\end{equation}
we can write
\begin{equation}
S_J^{(2)}(w_J)=\int d^3x dt\bar{N}_Ja_J^3\frac14 z_J^2
\left\{\frac12 \dot{w}_J^2-\frac12 \left(\frac{\partial_i w_J}{a_J}\right)^2\right\}
\,.
\label{Mukhanov action Jordan frame}
\end{equation}
Of course the second order action~\eqref{Mukhanov action Jordan frame} is also
obtained by starting with the original Jordan frame action
\eqref{JordanFrameAction} and doing the same steps as in the
Einstein frame derivation. That is, use the ADM decomposition,
solve for the constraint fields to first order, and write the
action for physical perturbations. In Ref.~\cite{Weenink:2010rr}
this was done without solving for the constraint fields, but
by decoupling the constraint degrees of freedom from the dynamical
degrees of freedom. The result for the scalar dynamical degree of
freedom agrees with~\eqref{Mukhanov action Jordan frame}.
This establishes the equivalence of frames at the level
of quadratic perturbations.

Note that when the action for perturbations is written
in terms of gauge invariant and frame independent perturbations,
there is no notion of separate frames. Although the Jordan frame
and Einstein frame actions \eqref{JordanFrameAction}
and \eqref{EinsteinFrameaction} look originally quite different
(nonminimal or minimal coupling), at the level of perturbations
the actions have a \textit{unique} form. Therefore it does not make
sense to talk about a preferred frame to describe perturbations.
One can choose to start with
an action for a certain nonminimally coupled scalar field, or for
another minimally coupled scalar field (related to the first
via the relation \eqref{EinsteinJordanFrameRelations})
in a modified potential, and both give equivalent results.
Of course, in the end one wants to express the results (such
as $n$-point functions) in terms of variables in the defining frame.

As an example, in Higgs inflation the defining frame is the Jordan frame,
where the scalar field $\Phi_J$
is identified with the Higgs field, the potential
$V_J$ is the Higgs potential and the field is coupled
to the scalar curvature via $F(\Phi_J)=1+\xi\Phi_J^2$. One can directly
compute perturbations in the Jordan frame, resulting in the manifestly
gauge invariant quadratic action \eqref{Mukhanov action Jordan frame},
but the derivation is difficult because of the nonminimal coupling.
It is easier to transform to the Einstein frame first, and then use
the standard result \eqref{Mukhanov gauge invariant action}. Of course
the scalar field in the Einstein frame is not the Higgs field, but
is related to the Higgs field via a field redefinition, just like the Einstein
frame metric. After finding the standard Einstein frame results
one can transform back to the Jordan frame (the defining frame) in order to obtain
\eqref{Mukhanov action Jordan frame}.

\subsection{Frame independence of cubic action}
\label{sec: Frame independence of cubic action}
In the previous section it was shown that the comoving curvature
perturbation in the Einstein frame coincides with the same perturbation
in the Jordan frame at first order. At second order however $w$ and
$w_J$ are related as
\begin{equation}
w=w_J+\left[\left(\frac{H_J}{\dot{\phi}_J}+\frac{\Omega'}{\bar{\Omega}}\right)
\frac{d^2\phi/d\phi_J^2}{d\phi/d\phi_J}
-\frac{\Omega^{\prime\prime}}{\bar{\Omega}}+\left(\frac{\Omega'}{\bar{\Omega}}\right)^2
\right]\varphi_J^2
\end{equation}
Thus the first order gauge invariant variable $w$ is not a convenient
variable for comparing Jordan frame results with Einstein frame results
at third order, since the variable itself is frame dependent.
Motivated by the results in the previous section, we could check if
a second order gauge invariant variable is frame independent,
in the sense that it has exactly the same form in the Jordan
and Einstein frames. Indeed, it can be shown that the curvature perturbation on uniform
field hypersurfaces, $W_{\zeta}$ from Eq.~\eqref{Wvarphigaugeinvariant},
coincides with the same gauge invariant variable in the Jordan frame,
\begin{align}
\nonumber W_{\zeta}&=
w-\left[\left(\frac{\ddot{\phi}}{H\dot{\phi}}-\frac{\dot{H}}{H^2}\right)
\frac{H^2}{\dot{\phi}^2}\varphi^2
+\frac{1}{\dot{\phi}}\dot{w}\varphi\right]\\
\nonumber &=w_J-\left[\left(\frac{\ddot{\phi}_J}{H_J\dot{\phi}_J}-\frac{\dot{H}_J}{H_J^2}\right)
\frac{H_J^2}{\dot{\phi}_J^2}\varphi_J^2
+\frac{1}{\dot{\phi}_J}\dot{w}_J\varphi_J\right]\\
&\equiv W_{\zeta,J}
\,,
\label{Equivalence Wvarphi}
\end{align}
where we did not write the terms with vectors, tensors or spatial derivatives.
The frame independence of $W_{\zeta}$ was shown explicitly
in~\cite{Sugiyama:2010zz}, but other, more general, proofs
exist as well~\cite{Koh:2005qp,Chiba:2008ia,Gong:2011qe}.
This means that we can directly find the third order action in the Jordan frame
from the third order Einstein frame action expressed in terms of $W_{\zeta}$,
which can be used, for example, to find $f_{\rm{NL}}$ for a nonminimally coupled
theory~\cite{Koh:2005qp,Koh:2005gy,Sugiyama:2010zz}. We only have to replace the background
Einstein frame quantities by corresponding Jordan frame quantities
using~\eqref{EinsteinJordan Background Relations}. Thus from \eqref{Cubic GI Action Wvarphi}
we straightforwardly find
\begin{align}
\nonumber \tilde{S}_{\rm{GI}}^{(3)}(W_{\zeta})=&\int d^3x dt\frac{\bar{N}_Ja_J^3}{8}\Biggl\{
\frac12 z_J^2 W_{\zeta,J}\left(\frac{\partial_i W_{\zeta,J}}{a_J}\right)^2
-\frac{3}{2}z_J^2 W_{\zeta,J}\dot{W}_{\zeta,J}^2
+\frac12 \frac{\bar{\Omega}z_J^2}{H_J+\frac{\dot{\bar{\Omega}}}{\bar{\Omega}}}\dot{W}_{\zeta,J}^3\\
&-\frac12 \left(3W_{\zeta,J}-\frac{\bar{\Omega}\dot{W}_{\zeta,J}}
{H_J+\frac{\dot{\bar{\Omega}}}{\bar{\Omega}}}\right)
\left[\frac{\partial_i\partial_j\psi_J}{a_J^2}\frac{\partial_i\partial_j\psi_J}{a_J^2}
-\left(\frac{\nabla^2\psi_J}{a_J^2}\right)^2\right]
+2\frac{\partial_i\psi_J}{a_J}\frac{\partial_iW_{\zeta,J}}{a_J}\frac{\nabla^2\psi_J}{a_J^2}
\Biggr\}
\,,
\label{Jordan: Cubic GI Action Wzeta}
\end{align}
where
\begin{equation}
\frac{\nabla^2\psi_J}{a_J^2}=-\frac{\nabla^2}{a_J^2}\frac{\bar{\Omega}W_{\zeta,J}}
{H_J+\frac{\dot{\bar{\Omega}}}{\bar{\Omega}}}+\frac12 z_J^2\dot{W}_{\zeta_J}
\,.
\end{equation}
The 3-point function in for a non-minimally coupled theory is found in a similar way.
Thus, one first transforms transforms to the Einstein frame where one can use
previously computed results for the 3-point function for $W_{\zeta}$, and this result
can be expressed in Jordan frame quantities by going back to the Jordan frame.

Of course we can also transform the partially integrated
Einstein frame action for $W_{\zeta}$
\eqref{Cubic GI Action Wvarphi after integrations},
that shows the separation between different orders in slow-roll,
to the Jordan frame.
After the frame transformations we find
\begin{align}
\nonumber \tilde{S}_{{\rm GI},J}^{(3)}=&\int d^3x dt \frac{\bar{N}_J a_J^3}{8} \Biggl\{
\frac{1}{2}\frac{z_J^4}{\bar{\Omega}^2}\left[-\frac12 \dot{W}_{\zeta,J}^2W_{\zeta,J}
-\frac12\left(\frac{\partial_i W_{\zeta,J}}{a_J}\right)^2W_{\zeta,J}
+\dot{W}_{\zeta,J}\left(\frac{\partial_i}{\nabla^2} \dot{W}_{\zeta,J}\right)\partial_i W_{\zeta,J}\right]\\
\nonumber &+\frac{1}{16}\frac{z_J^6}{\bar{\Omega}^4}\left[\dot{W}_{\zeta,J}^2W_{\zeta,J}
-\left(\frac{\partial_i\partial_j}{\nabla^2}\dot{W}_{\zeta,J}\right)
\left(\frac{\partial_i\partial_j}{\nabla^2}\dot{W}_{\zeta,J}\right)W_{\zeta,J}\right]
-\frac12 z_J^2 \dot{W}_{\zeta,J}W_{\zeta,J}^2
\left[\frac{\bar{\Omega}\left(z_J/\bar{\Omega}\right)^{\cdot}}
{z_J(H_J+\frac{\dot{\bar{\Omega}}}{\bar{\Omega}})}\right]^{\cdot}\\
&+\frac{1}{a_J^3}\frac{\delta S_J^{(2)}(w_J)}{\delta{w_J}}
\left[\frac{\bar{\Omega}}{4z_J (H_J+\frac{\dot{\bar{\Omega}}}{\bar{\Omega}})}
\left(z_J W_{\zeta,J}^2/\bar{\Omega}\right)^{\cdot}\right]
\Biggr\}
\,.
\label{Jordan: Cubic GI Action Wvarphi after integrations}
\end{align}
As explained in Sec.~\ref{sec: Uniqueness of gauge invariant action},
the boundary terms
are accounted for by performing a field redefinition
to a new gauge invariant variable
\begin{equation}
\tilde{W}_{\varphi,J}=W_{\zeta,J}+\frac{\bar{\Omega}}{4z_J (H_J+\frac{\dot{\bar{\Omega}}}{\bar{\Omega}})}
\left(z_J W_{\zeta,J}^2/\bar{\Omega}\right)^{\cdot}
\,.
\label{field redefinition Jordan frame}
\end{equation}
The 3-point function for $W_{\zeta,J}$ is then calculated from
the 3-point function for $\tilde{W}_{\varphi,J}$ plus disconnected parts in
that variable
\begin{align}
\nonumber \langle W_{\zeta,J}(\bvec{x}_1)& W_{\zeta,J}(\bvec{x}_2)
W_{\zeta,J}(\bvec{x}_3)\rangle = \langle \tilde{W}_{\varphi,J}(\bvec{x}_1)
\tilde{W}_{\varphi,J}(\bvec{x}_2)\tilde{W}_{\varphi,J}(\bvec{x}_3)\rangle\\
&+ \frac12\frac{\dot{z_J}-\frac{\dot{\bar{\Omega}}}{\bar{\Omega}}}
{z_J (H_J+\frac{\dot{\bar{\Omega}}}{\bar{\Omega}})}\left(
\langle \tilde{W}_{\varphi,J}(\bvec{x}_1)\tilde{W}_{\varphi,J}(\bvec{x}_2)\rangle
\langle \tilde{W}_{\varphi,J}(\bvec{x}_1)\tilde{W}_{\varphi,J}(\bvec{x}_3)\rangle
+\rm{sym}\right)
+\ldots
\,,
\label{bispectrum jordan}
\end{align}
where $z_J$ is defined in Eq.~\eqref{definition zj}
and terms of higher order in slow-roll have been neglected. The variable
$\tilde{W}_{\varphi,J}$ is of course the frame transformed $W_{\varphi}$,
which is seen most easily after transforming both sides of the
non-linear relation between gauge invariant variables
\eqref{difference gauge invariant variables}. $\tilde{W}_{\varphi,J}$
is however \textit{not} directly related to the field perturbation
on uniform curvature hypersurfaces in the Jordan frame, since
it does not reduce to $2 \frac{H_J}{\dot{\phi}_J}\varphi_J$  in the
gauge $\zeta_J=0$. In fact, the variable
\begin{equation}
W_{\varphi}\equiv w_J-
\left[\left(\frac{\ddot{\phi}_J}{H_J\dot{\phi}_J}-\frac{\dot{H}_J}{H_J^2}\right)(\zeta_J^2+w_J\zeta_J)
+\frac{1}{H}\dot{w}\zeta_J\right]
\,,
\end{equation}
is precisely the gauge invariant variable in the Jordan frame that does this.
The question is how to compute the 3-point function for the field perturbation
on uniform curvature hypersurfaces in the Jordan frame: what is the action for
$W_{\varphi,J}$? It can be shown that $W_{\varphi,J}$ is non-linearly
related to $\tilde{W}_{\varphi,J}$ as
\begin{equation}
\tilde{W}_{\varphi,J}=W_{\varphi,J}^2
+\frac14\left[\frac{\dot{z_J}-\frac{\dot{\bar{\Omega}}}{\bar{\Omega}}}
{z_J (H_J+\frac{\dot{\bar{\Omega}}}{\bar{\Omega}})}
-\frac{1}{\dot{\phi}_J}\left(\frac{\dot{\phi}_J}{H_J}\right)^{\cdot}\right]W_{\varphi,J}^2
\,,
\label{Jordan:nonlinear relation Wvarphis}
\end{equation}
where terms higher order in slow-roll have been neglected.
In the previous chapter \ref{sec: Uniqueness of gauge invariant action}
we have shown that the cubic actions for non-linearly related variables differ
only by boundary terms, which can give disconnected contributions
to the bispectrum. Thus the bispectrum for the field perturbation
on uniform curvature hypersurfaces in the Jordan frame contains a
connected part coming from the first lines of the action
\eqref{Jordan: Cubic GI Action Wvarphi after integrations}, plus
a disconnected part from the non-linear relation \eqref{Jordan:nonlinear relation Wvarphis}.
Alternatively, one could use the direct relation
\begin{equation}
W_{\varphi,J}=W_{\zeta,J}-\frac14\frac{1}{\dot{\phi}_J}\left(\frac{\dot{\phi}_J}{H_J}\right)^{\cdot}
W_{\zeta,J}^2
\,,
\label{Jordan:nonlinear relation WvarphiWzeta}
\end{equation}
by replacing all quantities by Jordan frame quantities in
\eqref{difference gauge invariant variables}. Then
one finds the connected part of the bispectrum for $W_{\varphi,J}$
from \eqref{Jordan: Cubic GI Action Wzeta}, and disconnected
pieces from \eqref{Jordan:nonlinear relation WvarphiWzeta}.

Now some words about the special situation when $\varphi_J=0$.
In that case the conformal factor $\Omega^2$ only has a background value
and $W_{\zeta}(\varphi=0)=\zeta_J=\zeta$.
Thus the cubic terms in $\zeta_J$ not only directly provide the gauge
invariant vertices, but can also be transformed to Einstein frame
vertices, and vice versa. The third order Jordan frame action for $\zeta_J$
was derived in~\cite{Qiu:2010dk}, and it was shown that it can be found
from the Einstein frame action~\cite{Maldacena:2002vr,Seery:2005wm,Chen:2006nt}
in Ref.~\cite{Kubota:2011re}. Moreover, one could imagine that at higher order
in perturbations one can construct a gauge invariant variable which
reduces to $\zeta_J$ in the gauge $\varphi_J=0$, just as we did before for $W_{\zeta}$
to second order. When $\varphi_J=0$ it now becomes
almost trivial to show that the curvature perturbation is invariant under frame transformations.
Thus, the curvature perturbation on uniform field hypersurfaces is frame independent
to all orders~\cite{Gong:2011qe}.

Finally a remark about previous results found in Higgs inflation.
Often computations are done in both frames, examples being
quantum corrections of the Higgs potential
\cite{Bezrukov:2008ej,Bezrukov:2009db,Barvinsky:2008ia,Barvinsky:2009fy,DeSimone:2008ei},
or the computations
of the cut-off scale for which the theory becomes nonperturbative
\cite{Lerner:2009na,Barbon:2009ya,Burgess:2010zq,Hertzberg:2010dc,Bezrukov:2010jz,Bezrukov:2011sz}.
The Einstein frame results are then compared to direct Jordan frame computations
by transforming them to the Jordan frame.
More often than not, transformed Einstein frame results do not exactly agree
with what is found in the Jordan frame. A recent example is
a calculation of the field dependent cut-off in \cite{Bezrukov:2011sz},
which appears different in one frame or the other. However,
the result of this section is that the Jordan frame
action can be found directly from the Einstein frame via a field transformation.
The most clear way to see this is that everything can be expressed
in frame independent variables.
Thus the cut-off scale should be the same whether you compute it directly
in the Jordan frame, or {\it via} transformed Einstein frame results.
The reason for the confusion and difference between results
obtained in different frames in the references above is
due to a non-covariant formalism, where the variables become
frame dependent. For example, in Einstein frame computations
often the nonperturbative cut-off scale
is found from expanding the non-polynomial potential
in powers of $\varphi$ and neglecting metric fluctuations. $\varphi$
is then a frame dependent variable. There is no confusion when
using frame independent variables: quantum corrections
or cut-off's computed directly in the Jordan frame
are exactly the same when first computed in the Einstein
frame and transformed back to the Jordan frame. The cut-off
can therefore in principle be computed directly from
\eqref{Jordan: Cubic GI Action Wzeta}, and from higher order
generalisation of that action.

\section{Summary and conclusion}
In this work we have focused on scalar perturbations of a scalar field
and the metric around an expanding background. These perturbations are
generally gauge dependent. It is possible to construct gauge invariant
variables by taking certain combinations of these perturbations.
At second order in coordinate transformations there are in principle
infinitely many gauge invariant variables, two
specific examples being the comoving curvature perturbation and
the field perturbation on uniform curvature hypersurfaces. These variables
are related in a non-linear way.

We have outlined the procedure for finding the gauge invariant actions
for these variables at third order, and we have shown explicitly the
gauge invariant cubic actions. Next we have demonstrated that, due to
the non-linear relation between these variables, the cubic actions appear
different, but actually their bulk actions are the same. They differ
by boundary terms, which generically give disconnected contributions
to the bispectrum.

This brings us to the aspect of uniqueness. Once
you pick a certain initial hypersurface, or equivalently choose
a specific gauge invariant variable, for example the comoving curvature
perturbation in a Gaussian state, then there is a unique action for
this variable that evolves the initial (Gaussian) state and creates non-Gaussianity
through the evolution. If the final hypersurface is different then the initial,
boundary terms must be taken into account which appear when switching
to a different gauge invariant variable associated with the final hypersurface.
We also commented on initial non-Gaussianity: although for one gauge invariant
variable the initial state is Gaussian, it is generically non-Gaussian
for another variable. Some of the final non-Gaussianity for a certain
variable therefore originates from some initial non-Gaussianity.

Finally we discussed different conformally related frames. It was shown that the cubic
action for gauge invariant perturbations in the Jordan frame can be obtained
directly from the Einstein frame action. The trick is to identify the variable
that has exactly the same form in either frame. The comoving curvature perturbation
is such a frame independent cosmological perturbation. Thus the bispectrum
for the comoving curvature perturbation can be found from the Einstein frame bispectrum
by transforming the result to the Jordan frame. The bispectrum for
another gauge invariant variable in the Jordan frame, such as the
field perturbation on uniform curvature hypersurfaces, is then found
from the non-linear relation between this variable and the comoving curvature perturbation.

Frame independent cosmological perturbations can be a very useful
tool in calculating quantum corrections for non-minimally coupled theories.
An example is Higgs inflation, where often different results are found
depending on which frame one uses for the computations due to non-covariance
of the perturbed actions. These differences can be attributed to both frame dependence and gauge
dependence of the formalism used. The issue is resolved when using frame independent, gauge invariant
variables, since then the Jordan and Einstein frame actions have exactly the same
form.
 
 The techniques used here can be quite straightforwardly generalised to
 compute higher order (quartic, etc.) gauge invariant tree level actions for
 scalar cosmological perturbations (needless to say, the choice of gauge
 invariant variables at cubic and higher orders is much richer), and it is the
 subject of future study. Higher order gauge invariant actions
 are the necessary tool for gauge invariant calculations of quantum corrections
 to cosmological observables, the simplest one being the one
 loop correction to the equal time two point function for scalar
 cosmological perturbations. This work represents a step towards establishing
 a fully covariant framework for calculating quantum field theoretic
 correlators in curved space time backgrounds.
 Cosmology is a field where such correlators may in fact be observable.
 The problem of observables
  in cosmology is one of the most important unsolved problems
 in theoretical cosmology\cite{WoodardTalk}.

\section{Acknowledgements}
We would like to thank Gerasimos Rigopoulos for useful discussions and comments.
This research was supported by the Dutch Foundation for
'Fundamenteel Onderzoek der Materie' (FOM) under the program
"Theoretical particle physics in the era of the LHC", program number FP 104.

\bibliography{Higgsinflation}{}

\providecommand{\href}[2]{#2}\begingroup\raggedright\begin{thebibliography}{10}

\bibitem{Maldacena:2002vr}
J.~M. Maldacena, {\it {Non-Gaussian features of primordial fluctuations in
  single field inflationary models}},  {\em JHEP} {\bf 05} (2003) 013,
  [\href{http://xxx.lanl.gov/abs/astro-ph/0210603}{{\tt astro-ph/0210603}}].

\bibitem{Seery:2005wm}
D.~Seery and J.~E. Lidsey, {\it {Primordial non-Gaussianities in single field
  inflation}},  {\em JCAP} {\bf 0506} (2005) 003,
  [\href{http://xxx.lanl.gov/abs/astro-ph/0503692}{{\tt astro-ph/0503692}}].

\bibitem{Chen:2006nt}
X.~Chen, M.-x. Huang, S.~Kachru, and G.~Shiu, {\it {Observational signatures
  and non-Gaussianities of general single field inflation}},  {\em JCAP} {\bf
  0701} (2007) 002, [\href{http://xxx.lanl.gov/abs/hep-th/0605045}{{\tt
  hep-th/0605045}}].

\bibitem{Bardeen:1980kt}
J.~M. Bardeen, {\it {Gauge Invariant Cosmological Perturbations}},  {\em Phys.
  Rev.} {\bf D22} (1980) 1882--1905.

\bibitem{Arnowitt:1962hi}
R.~L. Arnowitt, S.~Deser, and C.~W. Misner, {\it {The dynamics of general
  relativity}},  \href{http://xxx.lanl.gov/abs/gr-qc/0405109}{{\tt
  gr-qc/0405109}}.

\bibitem{Prokopec:2010be}
T.~Prokopec and G.~Rigopoulos, {\it {Path Integral for Inflationary
  Perturbations}},  {\em Phys. Rev.} {\bf D82} (2010) 023529,
  [\href{http://xxx.lanl.gov/abs/1004.0882}{{\tt arXiv:1004.0882}}].

\bibitem{Bruni:1996im}
M.~Bruni, S.~Matarrese, S.~Mollerach, and S.~Sonego, {\it {Perturbations of
  spacetime: Gauge transformations and gauge invariance at second order and
  beyond}},  {\em Class. Quant. Grav.} {\bf 14} (1997) 2585--2606,
  [\href{http://xxx.lanl.gov/abs/gr-qc/9609040}{{\tt gr-qc/9609040}}].

\bibitem{Malik:2008im}
K.~A. Malik and D.~Wands, {\it {Cosmological perturbations}},  {\em Phys.
  Rept.} {\bf 475} (2009) 1--51, [\href{http://xxx.lanl.gov/abs/0809.4944}{{\tt
  arXiv:0809.4944}}].

\bibitem{Noh:2004bc}
H.~Noh and J.-c. Hwang, {\it {Second-order perturbations of the Friedmann world
  model}},  {\em Phys. Rev.} {\bf D69} (2004) 104011.

\bibitem{Rigopoulos:2011eq}
G.~Rigopoulos, {\it {Gauge invariance and non-Gaussianity in Inflation}},  {\em
  Phys.Rev.} {\bf D84} (2011) 021301,
  [\href{http://xxx.lanl.gov/abs/1104.0292}{{\tt arXiv:1104.0292}}].

\bibitem{Malik:2003mv}
K.~A. Malik and D.~Wands, {\it {Evolution of second-order cosmological
  perturbations}},  {\em Class.Quant.Grav.} {\bf 21} (2004) L65--L72,
  [\href{http://xxx.lanl.gov/abs/astro-ph/0307055}{{\tt astro-ph/0307055}}].

\bibitem{Mukhanov:1990me}
V.~F. Mukhanov, H.~A. Feldman, and R.~H. Brandenberger, {\it {Theory of
  cosmological perturbations. Part 1. Classical perturbations. Part 2. Quantum
  theory of perturbations. Part 3. Extensions}},  {\em Phys. Rept.} {\bf 215}
  (1992) 203--333.

\bibitem{Weinberg:2005vy}
S.~Weinberg, {\it {Quantum contributions to cosmological correlations}},  {\em
  Phys.Rev.} {\bf D72} (2005) 043514,
  [\href{http://xxx.lanl.gov/abs/hep-th/0506236}{{\tt hep-th/0506236}}].

\bibitem{Arroja:2011yj}
F.~Arroja and T.~Tanaka, {\it {A note on the role of the boundary terms for the
  non-Gaussianity in general k-inflation}},  {\em JCAP} {\bf 1105} (2011) 005,
  [\href{http://xxx.lanl.gov/abs/1103.1102}{{\tt arXiv:1103.1102}}].

\bibitem{Burrage:2011hd}
C.~Burrage, R.~H. Ribeiro, and D.~Seery, {\it {Large slow-roll corrections to
  the bispectrum of noncanonical inflation}},  {\em JCAP} {\bf 1107} (2011)
  032, [\href{http://xxx.lanl.gov/abs/1103.4126}{{\tt arXiv:1103.4126}}].

\bibitem{Salopek:1988qh}
D.~S. Salopek, J.~R. Bond, and J.~M. Bardeen, {\it {Designing Density
  Fluctuation Spectra in Inflation}},  {\em Phys. Rev.} {\bf D40} (1989) 1753.

\bibitem{Bezrukov:2007ep}
F.~L. Bezrukov and M.~Shaposhnikov, {\it {The Standard Model Higgs boson as the
  inflaton}},  {\em Phys. Lett.} {\bf B659} (2008) 703--706,
  [\href{http://xxx.lanl.gov/abs/0710.3755}{{\tt arXiv:0710.3755}}].

\bibitem{Makino:1991sg}
N.~Makino and M.~Sasaki, {\it {The Density perturbation in the chaotic
  inflation with nonminimal coupling}},  {\em Prog. Theor. Phys.} {\bf 86}
  (1991) 103--118.

\bibitem{Fakir:1992cg}
R.~Fakir, S.~Habib, and W.~Unruh, {\it {Cosmological density perturbations with
  modified gravity}},  {\em Astrophys. J.} {\bf 394} (1992) 396.

\bibitem{Hwang:1996np}
J.-c. Hwang, {\it {Cosmological perturbations in generalized gravity theories:
  Conformal transformation}},  {\em Class. Quant. Grav.} {\bf 14} (1997)
  1981--1991, [\href{http://xxx.lanl.gov/abs/gr-qc/9605024}{{\tt
  gr-qc/9605024}}].

\bibitem{Weenink:2010rr}
J.~Weenink and T.~Prokopec, {\it {Gauge invariant cosmological perturbations
  for the nonminimally coupled inflaton field}},  {\em Phys.Rev.} {\bf D82}
  (2010) 123510, [\href{http://xxx.lanl.gov/abs/1007.2133}{{\tt
  arXiv:1007.2133}}].

\bibitem{Sugiyama:2010zz}
N.~Sugiyama and T.~Futamase, {\it {Non-Gaussianity generated in the
  inflationary scenario with nonminimally coupled inflaton field}},  {\em
  Phys.Rev.} {\bf D81} (2010) 023504.

\bibitem{Koh:2005qp}
S.~Koh, {\it {Non-gaussianity in nonminimally coupled scalar field theory}},
  {\em J.Korean Phys.Soc.} {\bf 49} (2006) S787--S790,
  [\href{http://xxx.lanl.gov/abs/astro-ph/0510030}{{\tt astro-ph/0510030}}].

\bibitem{Chiba:2008ia}
T.~Chiba and M.~Yamaguchi, {\it {Extended Slow-Roll Conditions and Rapid-Roll
  Conditions}},  {\em JCAP} {\bf 0810} (2008) 021,
  [\href{http://xxx.lanl.gov/abs/0807.4965}{{\tt arXiv:0807.4965}}].

\bibitem{Gong:2011qe}
J.-O. Gong, J.-c. Hwang, W.-I. Park, M.~Sasaki, and Y.-S. Song, {\it {Conformal
  invariance of curvature perturbation}},  {\em JCAP} {\bf 1109} (2011) 023,
  [\href{http://xxx.lanl.gov/abs/1107.1840}{{\tt arXiv:1107.1840}}].

\bibitem{Koh:2005gy}
S.~Koh, S.~P. Kim, and D.~J. Song, {\it {Nonlinear evolutions and
  non-Gaussianity in generalized gravity}},  {\em Phys.Rev.} {\bf D71} (2005)
  123511, [\href{http://xxx.lanl.gov/abs/astro-ph/0501401}{{\tt
  astro-ph/0501401}}].

\bibitem{Qiu:2010dk}
T.~Qiu and K.-C. Yang, {\it {Non-Gaussianities of Single Field Inflation with
  Non-minimal Coupling}},  {\em Phys.Rev.} {\bf D83} (2011) 084022,
  [\href{http://xxx.lanl.gov/abs/1012.1697}{{\tt arXiv:1012.1697}}].

\bibitem{Kubota:2011re}
T.~Kubota, N.~Misumi, W.~Naylor, and N.~Okuda, {\it {The Conformal
  Transformation in General Single Field Inflation with Non-Minimal Coupling}},
   {\em JCAP} {\bf 1202} (2012) 034,
  [\href{http://xxx.lanl.gov/abs/1112.5233}{{\tt arXiv:1112.5233}}].

\bibitem{Bezrukov:2008ej}
F.~L. Bezrukov, A.~Magnin, and M.~Shaposhnikov, {\it {Standard Model Higgs
  boson mass from inflation}},  {\em Phys. Lett.} {\bf B675} (2009) 88--92,
  [\href{http://xxx.lanl.gov/abs/0812.4950}{{\tt arXiv:0812.4950}}].

\bibitem{Bezrukov:2009db}
F.~Bezrukov and M.~Shaposhnikov, {\it {Standard Model Higgs boson mass from
  inflation: two loop analysis}},  {\em JHEP} {\bf 07} (2009) 089,
  [\href{http://xxx.lanl.gov/abs/0904.1537}{{\tt arXiv:0904.1537}}].

\bibitem{Barvinsky:2008ia}
A.~O. Barvinsky, A.~Y. Kamenshchik, and A.~A. Starobinsky, {\it {Inflation
  scenario via the Standard Model Higgs boson and LHC}},  {\em JCAP} {\bf 0811}
  (2008) 021, [\href{http://xxx.lanl.gov/abs/0809.2104}{{\tt
  arXiv:0809.2104}}].

\bibitem{Barvinsky:2009fy}
A.~O. Barvinsky, A.~Y. Kamenshchik, C.~Kiefer, A.~A. Starobinsky, and
  C.~Steinwachs, {\it {Asymptotic freedom in inflationary cosmology with a non-
  minimally coupled Higgs field}},  {\em JCAP} {\bf 0912} (2009) 003,
  [\href{http://xxx.lanl.gov/abs/0904.1698}{{\tt arXiv:0904.1698}}].

\bibitem{DeSimone:2008ei}
A.~De~Simone, M.~P. Hertzberg, and F.~Wilczek, {\it {Running Inflation in the
  Standard Model}},  {\em Phys. Lett.} {\bf B678} (2009) 1--8,
  [\href{http://xxx.lanl.gov/abs/0812.4946}{{\tt arXiv:0812.4946}}].

\bibitem{Lerner:2009na}
R.~N. Lerner and J.~McDonald, {\it {Higgs Inflation and Naturalness}},  {\em
  JCAP} {\bf 1004} (2010) 015, [\href{http://xxx.lanl.gov/abs/0912.5463}{{\tt
  arXiv:0912.5463}}].

\bibitem{Barbon:2009ya}
J.~L.~F. Barbon and J.~R. Espinosa, {\it {On the Naturalness of Higgs
  Inflation}},  {\em Phys. Rev.} {\bf D79} (2009) 081302,
  [\href{http://xxx.lanl.gov/abs/0903.0355}{{\tt arXiv:0903.0355}}].

\bibitem{Burgess:2010zq}
C.~P. Burgess, H.~M. Lee, and M.~Trott, {\it {Comment on Higgs Inflation and
  Naturalness}},  {\em JHEP} {\bf 07} (2010) 007,
  [\href{http://xxx.lanl.gov/abs/1002.2730}{{\tt arXiv:1002.2730}}].

\bibitem{Hertzberg:2010dc}
M.~P. Hertzberg, {\it {On Inflation with Non-minimal Coupling}},  {\em JHEP}
  {\bf 1011} (2010) 023, [\href{http://xxx.lanl.gov/abs/1002.2995}{{\tt
  arXiv:1002.2995}}].

\bibitem{Bezrukov:2010jz}
F.~Bezrukov, A.~Magnin, M.~Shaposhnikov, and S.~Sibiryakov, {\it {Higgs
  inflation: consistency and generalisations}},  {\em JHEP} {\bf 1101} (2011)
  016.

\bibitem{Bezrukov:2011sz}
F.~Bezrukov, D.~Gorbunov, and M.~Shaposhnikov, {\it {Late and early time
  phenomenology of Higgs-dependent cutoff}},  {\em JCAP} {\bf 1110} (2011) 001,
  [\href{http://xxx.lanl.gov/abs/1106.5019}{{\tt arXiv:1106.5019}}].

\bibitem{WoodardTalk}
R.~P. Woodard ,~Talk at the CERN 2011 workshop on "Quantum Gravity: from UV to
  IR".

\end{thebibliography}\endgroup
\bibliographystyle{JHEP}

\end{document}